\documentclass[fleqn,usenatbib]{mnras}

\usepackage{newtxtext,newtxmath}

\usepackage[T1]{fontenc}

\DeclareRobustCommand{\VAN}[3]{#2}
\let\VANthebibliography\thebibliography
\def\thebibliography{\DeclareRobustCommand{\VAN}[3]{##3}\VANthebibliography}

\usepackage{graphicx}	
\usepackage{amsmath}	
\usepackage{soul}
\usepackage{bm}

\usepackage[encapsulated]{CJK}
\usepackage{ucs}
\usepackage[utf8]{inputenc}
\usepackage{xcolor}

\newcommand{\tctext}[1]{\begin{CJK}{UTF8}{bkai}#1\ignorespacesafterend\end{CJK}}

\newcommand{\rn}[1]{\textcolor{magenta}{#1}}

\title[Discs around flybys]{Formation of misaligned second-generation discs through flyby encounters}

\author[Smallwood et al.]{Jeremy L. Smallwood,$^1$\thanks{E-mail: jlsmallwood@asiaa.sinica.edu.tw}
Rebecca Nealon,$^{2,3}$ Nicol\'as Cuello$^{4}$,
Ruobing Dong (\tctext{董若冰})$^{5,1}$
and
\newauthor
Richard A. Booth$^{6}$
\\
$^1$Institute of Astronomy and Astrophysics, Academia Sinica, Taipei 10617, R.O.C.\\
$^2$Centre for Exoplanets and Habitability, University of Warwick, Coventry CV4 7AL, UK\\
$^3$Department of Physics, University of Warwick, Coventry CV4 7AL, UK\\
$^4$Univ. Grenoble Alpes, CNRS, IPAG, 38000 Grenoble, France\\
$^5$Department of Physics and Astronomy, University of Victoria, Victoria, BC V8P 5C2, Canada\\
$^6$School of Physics and Astronomy, University of Leeds, Leeds, LS2 9JT, UK\\
}

\date{Accepted XXX. Received YYY; in original form ZZZ}

\pubyear{2023}

\begin{document}
\label{firstpage}
\pagerange{\pageref{firstpage}--\pageref{lastpage}}
\maketitle

\begin{abstract}
 Observations reveal protoplanetary discs being perturbed by flyby candidates. We simulate a scenario where an unbound perturber, i.e., a flyby, undergoes an inclined grazing encounter, capturing material and forming a second-generation protoplanetary disc. We run $N$--body and three-dimensional hydrodynamical simulations of a parabolic flyby grazing a particle disc and a gas-rich protoplanetary disc, respectively.  In both our $N$--body and hydrodynamic simulations, we find that the captured, second-generation disc forms at a tilt twice the initial flyby tilt. This relationship  is robust to variations in the flyby's tilt, position angle, periastron, and mass. We extend this concept by also simulating the case where the flyby has a disc of material prior to the encounter but we do not find the same trend. An inclined disc with respect to the primary disc around a misaligned flyby is tilted by a few degrees, remaining close to its initial disc tilt. Therefore, if a disc is present around the flyby before the encounter, the disc may not tilt up to twice the perturber tilt depending on the balance between the angular momentum of the circumsecondary disc and captured particles. In the case where the perturber has no initial disc, analyzing the  orientation of these second-generation discs can give information about the orbital properties of the flyby encounter.
\end{abstract}




\section{Introduction}
Recent observations of protoplanetary discs reveal disc substructures, such as rings, gaps, and spirals \citep{Andrews2020,vanderMarel2021}. Disc substructures can be excited from either bound or unbound companions, suggesting that these substructures can be used as a signpost for planet formation \citep{Grady1999,Grady2013,Muto2012,Wagner2015,Monnier2019,Garufi2020,MuroArena2020}.  Stars born in dense stellar clusters are subject to stellar flyby events \citep{Pfalzner2013}, where a companion on an unbound orbit can perturb protoplanetary discs \citep{Clarke1993,Cuello2023}. Studying the long-term effects of a flyby encounter on the protoplanetary disc structure can shed light on understanding observations.

A perturber on a flyby or unbound orbit is defined as having a single periastron passage within $1000\, \rm au$. The probability of stellar flyby events is enhanced in dense stellar clusters, where the chance of stellar encounters is high \citep{Hillenbrand1997,Carpenter2000,Lada2003,Porras2003}. From the works of \cite{Pfalzner2013}, and \cite{Winter2018a}, stellar flybys encounter a solar-type star within the first million years of stellar evolution at a probability of $30\%$ for a background stellar density that is larger than in Taurus. Recently, \cite{Pfalzner2021} found that the frequency of close flybys in low-mass clusters is underestimated and that low-mass clusters should contain $10\%-15\%$ of discs smaller than $30\,\rm au$ truncated by flybys. The hydrodynamical studies of star formation from dense stellar clusters by \cite{Bate2018} reported that most stellar encounters occur with the first Myr of stellar evolution, consistent with previous works. Parabolic orbit encounters are found to be more probable than hyperbolic orbits \cite[see Fig.7 from][]{Pfalzner2013}.  The lifetime of gaseous protoplanetary discs is estimated to be $1-10\, \rm Myr$ \citep{Haisch2001,Hernandez2007,Hernandez2008,Mamajek2009,Ribas2015}. Therefore, flyby events have the potential to perturb and shape protoplanetary discs \citep{Cuello2019,Cuello2020,Jimenez-Torres2020,Menard2020}. For example, unbound encounters can truncate protoplanetary discs, which can influence the total size and occurrence rate of planetary systems \cite[e.g.,][]{Scally2001,Adams2006,Olczak2006,Steinhausen2014,Rosotti2014,PortegiesZwart2016,Vincke2016,ConchaRamirez2019,Jimenez-Torres2020,Concharamirez2021}. Stellar flybys can enhance photoevaporation of protoplanetary discs, which can ultimately decrease the gaseous disc lifetime \citep{Dai2018,Winter2018a}. 

There are several observed flyby candidates that are undergoing interactions with protoplanetary discs, such as RW Aur \citep{Cabrit2006,Dai2015,Rodriguez2018}, AS 205 \citep{Kurtovic2018}, HV Tau and Do Tau \citep{Winter2018b}, FU Ori \citep{Beck2012,Takami2018,Perez2020,Borchert2022a,Borchert2022b},  Z CMa \citep{Takami2018,Dong2022}, UX Tau \citep{Menard2020}, and Sgr C \citep{Lu2022}. The systems V2775 Ori \citep{Zurlo2017} and V1647 Ori \citep{Principe2018} are highly speculative to be flyby encounters.  For a recent review on flyby's shaping protoplanetary discs, see \cite{Cuello2023}. 

When the perturber approaches periastron passage,  tidal effects by the perturber excites the formation of spirals and potentially disc fragmentation \citep{Ostriker1994,Pfalzner2003,Shen2010,Thies2010,Smallwood2023}. External unbound companions will excite spiral density waves at Lindblad, and corotation resonances \cite[e.g.,][]{Lin1993}. If the unbound companion is an external star, it exerts a strong tidal force where its Roche lobe can reach beyond the location of most of these resonances. Furthermore, flyby events can warp the primary disc for a range of perturber inclinations and periastron distances \citep{Clarke1993,Ostriker1994,Terquem1996,Bhandare2016,Xiang-Gruess2016}. Aside from spiral formation, long bridges of material are linked from the primary disc to the intruding flyby \citep{Cuello2019,Cuello2020}.   Warps and misalignments are typical in the primary disc and are observable in moment one maps \citep{Cuello2020}. Broken protoplanetary discs can have large mutual misalignments between the inner and outer gas rings generated by a flyby scenario \citep{Nealon2020}. 

\cite{Clarke1993} demonstrated that a prograde, coplanar parabolic flyby encounter stripped material off the protoplanetary disc, and the perturber captured a portion of the stripped material. Perturbers on hyperbolic trajectories ($e>1$) have a higher angular velocity during periapsis, leaving a lesser mark on the primary disc structure \cite[e.g.,][]{Winter2018b}, and are less efficient in capturing material compared to parabolic encounters \cite[e.g.,][]{Larwood1997,Pfalzner2005b,Breslau2017}. Despite knowing that material can be captured during a flyby encounter, the relationship between the inclination of the perturber and the captured material has not been investigated fully. \cite{Jikova2016} examined the distribution of  captured material during a flyby encounter through $N$--body simulations, however, they did not consider hydrodynamical discs.  After the passage of the perturber has already occurred, we can still observe the second generation disc. Therefore, if there is a relationship between the captured material and the flyby, we can reconstruct the orbit of the flyby and shed light on the many flyby candidate observations. One observational example is UX Tau, where the disc around UX Tau C is thought to be captured during the encounter \citep{Menard2020}. 

In this work, we focus on the transfer of material from the primary protoplanetary disc to the unbound perturber, which forms a second-generation disc. We run 3-dimensional $N$--body and hydrodynamical simulations of a parabolic flyby interacting with a protoplanetary disc,  tracking the formation and evolution of disc material around the flyby.  We find there is a strong relationship between the inclination of the captured material and the initial tilt of the perturber.  By measuring the mutual inclination of the two discs and comparing them to observations, we can reconstruct the orbit of the observed flyby candidate, deducing whether or not flyby candidates are indeed on unbound orbits.   The layout of the paper is as follows. Section~\ref{sec::methods} describes the numerical setup routines for our $N$--body and hydrodynamical simulations to model a parabolic encounter interacting with a circumprimary disc. In Sections~\ref{sec::Nbody_results} and~\ref{sec::hydro_results}, we report the results of our $N$--body and hydrodynamical simulations, respectively. Section~\ref{sec::two_discs} shows hydrodynamical results of two interacting protoplanetary discs on parabolic orbits. Section~\ref{sec::analytics} gives an analytical framework on how particles are captured during a flyby encounter. In Section~\ref{sec::discussion}, we discuss how our results apply to observations of flyby candidate systems. Finally, we give a conclusion in Section~\ref{sec::summary}.


\section{methods}
\label{sec::methods}
 We conduct two types of simulations. First, we consider a flyby using an $N$--body code which does not take into account any pressure or viscous effects. Second we confirm and expand on these results using hydrodynamic simulations.  For the hydrodynamical simulations, we simulate a bound, parabolic, and hyperbolic encounter to test how the relative velocity between the perturber and disc affects the orientation of the captured disc around the perturber. For the $N$--body simulations, we only simulate a parabolic encounter. Here we detail the important parameters for all of our simulations.

\subsection{Parabolic orbit setup}
We describe the setup of an unbound perturber that gravitationally influences the protoplanetary disc around the primary star. We simulate strictly parabolic encounters ($e \approx 1$), which induce the largest star-to-disc angular momentum transfer and produce the most prominent substructures in the disc \citep{Vincke2016,Winter2018b,Cuello2019,Cuello2020}.

\begin{figure*} 
\centering
\includegraphics[width=2\columnwidth]{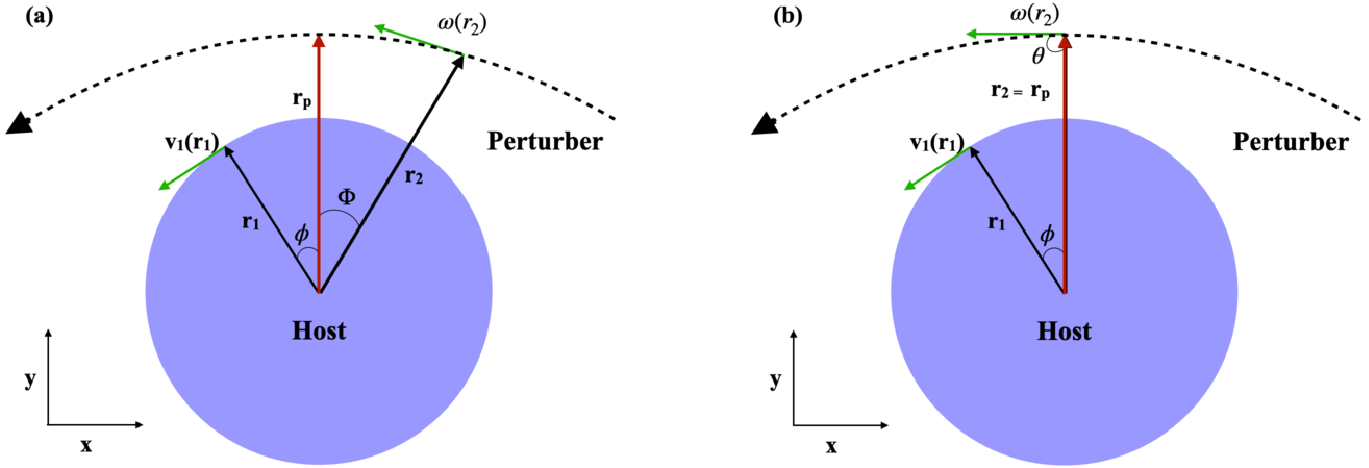}
\centering
\caption{A schematic view of an accretion disc (the host) encountering a perturber on a parabolic orbit (dotted-black curve). The left panel shows before closest approach and the right panel the instant of closest approach, used to define the terms in Equation~\ref{eq::rp_and_theta}. The position vectors are given in black, with the exception of the periastron position vector which is given in red. The velocity vectors are given in green.   }
\label{fig::flyby_orbit}
\end{figure*}

 We use the same orbital setup for our $N$--body and hydrodynamical simulations.  In this work we denote the host and flyby with subscripts "1" and "2", respectively. A schematic of a perturber on a parabolic orbit encountering an accretion disc is given in Fig.~\ref{fig::flyby_orbit}.  We model coplanar and inclined parabolic trajectories with the radial distance, $r_2$, described by 
\begin{equation}
    r_2 = \frac{2r_{\rm p}}{1 + \sin\theta},
    \label{eq::rp_and_theta}
\end{equation}
\citep{Bate1971} where $r_{\rm p}$ is the periastron distance, and  $\theta$ is the angle between periastron position vector and velocity vector. The periastron passage occurs at $\theta = +\pi / 2$, where the velocity vector is perpendicular to the periastron position vector (see the right panel in Fig.~\ref{fig::flyby_orbit}). The angular speed as a function of $r_2$ is then
\begin{align}
    \omega(r_2) &=
    \sqrt{\frac{2G(M_1 + M_2)}{r_{\rm p}^3}}
    \left(\frac{r_{\rm p}}{r_2}\right)^2\\&=
    \left(0.18\,^\circ\textrm{yr}^{-1}\right)
    \left(\frac{M_1 + M_2}{M_{\sun}}\right)^{1/2}
    \left(\frac{r_\mathrm{p}}{200\,\textrm{au}}\right)^{-3/2}
    \left(\frac{r_{\rm p}}{r_2}\right)^2,
    \label{eq::omega}
\end{align}
\citep{Bate1971} where $M_1$ and $M_2$ are the masses of the primary and flyby, respectively,  $G$ is the gravitational constant. The relationship between the flyby separation, $r_2$, and time $t$ is given by
\begin{equation}
    \left(\frac{r_2}{r_{\rm p}} + 2 \right) \sqrt{\frac{r_2}{r_{\rm p}} - 1} = \frac{3|t - t_\mathrm{p}|}{2} \sqrt{\frac{2G (M_1 + M_2)}{r_{\rm p}^3}},
    \label{eq::rp}
\end{equation}
where $r_2 = r_\mathrm{p}$ when $t = t_\mathrm{p}$. 

For a coplanar parabolic orbit, the perturber lies in the $x$--$y$ plane,  arrives initially from the negative $y$, positive x direction, and leaves towards the negative $y$, negative x direction. When we incline the orbit by an arbitrary amount, we rotate the orbit clock-wise about the $y$--axis. Therefore, all coplanar and inclined models will have the same perturber periastron ($x$,$y$,$z$) coordinate centered on the host star, unless the position angle of the orbit is non-zero.

\subsection{\texorpdfstring{$N$}{}--body simulation setup}
We model a perturber on a parabolic orbit and a circumprimary disc of particles using the {\sc whfast} integrator, which is a second-order symplectic Wisdom Holman integrator with 11th-order symplectic correctors in the $N$-body simulation package, {\sc rebound} \citep{Rein2015}. We construct a disc of $10,000$ test particles around the primary star, with an inner disc radius $r_{\rm in} = 10\, \rm au$, and outer disc radius $r_{\rm out} = 100\, \rm au$. The test particles are initially on circular orbits and coplanar with respect to the $x$-$y$ plane. The centre star has a mass $M_1 = 1\, \rm M_{\odot}$, and the perturber's mass is also set to $M_2 = 1\, \rm M_{\odot}$.  The perturber's periastron distance is set to $r_{\rm p} = 200\, \rm au$, with an initial separation $r_0 = 500\, \rm au$. We model various simulations where  parabolic orbit is tilted by $i_0 = 0^\circ$, $15^\circ$, $30^\circ$, $45^\circ$, and $60^\circ$, see Table~\ref{table::n_body}.  For each particle in the simulation, we determine whether it is bound to the primary or secondary star by calculating the specific energies (kinetic plus potential). When the specific energies of the particles are negative, they are considered bound,
and we then calculate the particle parameters (i.e., separation, eccentricity, and inclination) with respect to its bound companion.  The inclination is measured by calculating the angle between the particle's angular momentum vector and the $z$-axis.  
Observationally, it is more useful to analyze the inclination from the $z$--axis as it indicates the angle by which the orbits have been inclined with respect to the initial state.

\begin{table}
	\centering
	\caption{A summary of the $N$--body simulations. The simulation ID is given in the first column. The tilt of the perturber is given in the second column, and the  average tilt of the captured particles around the perturber with the standard error} is given in the last column.
	\label{table::n_body}
	\begin{tabular}{ccc} 
		\hline
		Simulation  & $i_2/^\circ$  & $i_{\rm disc,2}/^\circ$ \\
		\hline
		N0   & $0$ & $\sim 0 \pm 0.0$ \\
		N15 & $15$ & $\sim 43 \pm 0.521 $ \\
		N30  & $30$ &  $\sim 63 \pm 0.525$ \\
		N45 & $45$ & $\sim 81 \pm 0.667 $ \\
		N60 & $60$ & $\sim 113 \pm 3.485 $ \\
	    \hline
	\end{tabular}
\end{table}

\begin{table*}
	\centering
	\caption{The setup of the SPH simulations. The simulation ID is given in the first column.  The remaining columns lists the mass of the perturber, $M_2$, the distance of closest approach, $r_{\rm p}$, the initial tilt of the flyby orbit, $i_2$, the position angle of the flyby orbit, $\rm PA_2$, the number of particles, and the tilt of the perturber disc, $i_{\rm disc,2}$,  along with the standard error.}
	\label{table::setup}
	\begin{tabular}{cccccccc} 
		\hline
		Simulation ID &  $M_2/\rm M_\odot$ & $e_2$  & $r_{\rm p}/\rm au$ & $i_2/^\circ$ & $\rm PA_2/^\circ$ & \# of particles  & $i_{\rm disc,2}/^\circ$    \\
		\hline
		H0 & $0.2$ & $1$ & $100$ & $0$ & $0$& $5\times 10^5$  & $\sim 0 \pm 0.008$\\
		H45 & $0.2$ & $1$ & $100$ & $45$ & $0$& $5\times 10^5$  & $\sim 90 \pm 0.073$ \\
		H15 & $0.2$ & $1$ & $100$ & $15$ & $0$& $5\times 10^5$  & $\sim 30 \pm 0.066$  \\
		H30 & $0.2$ & $1$ & $100$ & $30$ & $0$& $5\times 10^5$  & $\sim 60 \pm 0.060$ \\
		H45HR & $0.2$ & $1$ & $100$ & $45$ & $0$& $4\times 10^6$  & $\sim 90 \pm 0.041$  \\
		H60 & $0.2$ & $1$ & $100$ & $60$ & $0$& $5\times 10^5$  & $\sim 120 \pm 0.098$ \\
		\hline
		H45PA30 & $0.2$ & $1$ & $100$ & $45$ & $30$& $5\times 10^5$  & $\sim 90 \pm 0.032$  \\
		H45PA60 & $0.2$ & $1$ & $100$ & $45$ & $60$& $5\times 10^5$  & $\sim 90 \pm 1.634$  \\
		H45PA90 & $0.2$ & $1$ & $100$ & $45$ & $90$& $5\times 10^5$  & $\sim 90 \pm 1.850$  \\
		\hline
		H45R120 & $0.2$ & $1$ & $120$ & $45$ & $0$& $5\times 10^5$  & $\sim 90 \pm 0.040$  \\
		H45R80 & $0.2$ & $1$ & $80$ & $45$ & $0$& $5\times 10^5$  & $\sim 90 \pm 0.029$  \\
		\hline
		H45M1 & $1$ & $1$ & $100$ & $45$ & $0$& $5\times 10^5$   & $\sim 90 \pm 0.039$ \\
            \hline
            H45p3 & $0.2$ & $0.3$ & $100$ & $45$ & $0$& $5\times 10^5$  & $\sim 92 \pm 0.164$ \\
            H45p5 & $0.2$ & $0.7$ & $100$ & $45$ & $0$& $5\times 10^5$  & $\sim 90 \pm 0.492$ \\
		
		\hline
	\end{tabular}
\end{table*}

\subsection{Hydrodynamical simulation setup}
We simulate a primary star surrounded by a gaseous protoplanetary disc and a parabolic flyby encounter using the 3-dimensional smoothed particle hydrodynamics code {\sc phantom} \citep{Price2018}. {\sc phantom} has been extensively tested to simulate unbound encounters \citep{Cuello2019,Cuello2020,Nealon2020,Menard2020,Borchert2022a,Borchert2022b,Smallwood2023}. The code can model an assortment of parabolic orbit configurations such that the system's angular momentum is conserved with the same accuracy order as the time-stepping scheme. We only report encounters that result in a disc around the perturber. 

\subsubsection{Primary star and protoplanetary disc setup}
\label{sec::primarydisc_setup}
We set up a gas-only protoplanetary disc around a generic solar-type star that is initially coplanar to the spin-axis of the star, assumed to be the $z$--axis. We simulate the hydrodynamical disc in the bending wave regime, such that the disc aspect ratio $H/r$ is larger than the \cite{Shakura1973} viscosity coefficient $\alpha_{\rm SS}$.
The warp induced by the unbound perturber will propagate as a pressure wave with speed $\sim c_s/2$ \citep{Papaloizou1983,Papaloizou1995}, where $c_{\rm s}$ is the sound speed. The hydrodynamical disc is modelled as a flat disc with $500,000$ Lagrangian particles with a total disc mass of $0.001\, \rm M_{\odot}$. We include one higher resolution simulation with $4\times 10^6$ particles for a resolution study. During periastron passage of the flyby, the low disc mass ensures that there is negligible gravitational effect imparted onto the flyby from the disc  and we can safely ignore the effect of disc self-gravity.  The mass of the  primary star is set to $M_1 = 1\, \rm M_{\odot}$.  We set the inner disc radius to $r_{\rm in} = 10\, \rm au$ and the outer radius is $r_{\rm out} = 100\, \rm au$. The primary star has an accretion radius of $r_{\rm acc,1} = 10\, \rm au$. We purposefully make the accretion radius equivalent to the initial inner edge of the disc to speed up computational time with not having to resolve close-in particle orbits. The accretion radius is a hard boundary such that any Lagrangian particles that penetrate the boundary are considered accreted, and the particle's mass, angular momentum and linear momentum are deposited onto the sink.

The disc surface density profile is initially a power law distribution given by
 \begin{equation}
     \Sigma(R) = \Sigma_0 \bigg( \frac{r}{r_{\rm in}} \bigg)^{-p},
     \label{eq::sigma}
 \end{equation}
where $\Sigma_0 = 7.00\, \rm g/cm^2$ is the density normalization, $r$ is the radial distance in the disc, and $p$ is the power law index. We set $p=1.5$, 
and the total disc mass defines the density normalization.   Previous hydrodynamics simulations of flyby-disc interactions used a radial surface density profile of $p=1$ to match observed disc profiles \cite[e.g.,][]{Cuello2019,Cuello2020}, which initially loads more material in the outer disc regions compared to $p=1.5$. Since we select a low disc mass, the dynamical behaviour of the disc material during the encounter does not sensitively depend on the initial surface density profile. We use a locally isothermal equation-of-state with a disc thickness that is scaled with radius as  
\begin{equation}
    H = \frac{c_{\rm s}}{\Omega} \propto r^{3/2-q}, 
 \end{equation}
where $\Omega = \sqrt{GM/r^3}$. The initial disc aspect ratio is $H/r = 0.05$ at $r_{\rm in}$. The \cite{Shakura1973} viscosity prescription, denoted as $\alpha_{\rm SS}$, is given by 
\begin{equation}
    \nu = \alpha_{\rm SS} c_{\rm s} H,
\end{equation}
where $\nu$ is the kinematic viscosity. To calculate $\alpha_{\rm SS}$, we follow the details given in \cite{Lodato2010}, such that
\begin{equation}
\alpha_{\rm SS} \approx \frac{\alpha_{\rm AV}}{10}\frac{\langle h \rangle}{H},
\end{equation}
where $\langle h \rangle$ is the mean smoothing length of particles in a cylindrical ring at a given radius \citep{Lodato2010}.  In this work, we set $\alpha_{\rm SS} = 0.005$, which translates to an artificial viscosity of $\alpha_{\rm AV} = 0.1260$ ($0.2713$ for the high-resolution simulation) 
 \cite[see][for details]{Meru2012}. We note that the $\alpha_{\rm AV}$ is always higher than the suggested limit from \citet{Meru2012}. To prevent particle-particle penetration in the high Mach number regime, we include a term, $\beta_{\rm AV}$ \cite[e.g.,][]{Monaghan1989}. Traditionally,  $\beta_{\rm AV} = 2.0$ \citep{Lodato2007,Price2018}. 
The disc is resolved with a shell-averaged smoothing length per scale height of $\langle h \rangle/H \approx 0.5$ and $\langle h \rangle/H \approx 0.25$ for our high-resolution simulation. 

To more accurately simulate the formation and development of  discs around an unbound companion, we adopt the locally isothermal equation of state of \cite{Farris2014} and set the sound speed $c_{\rm s}$ to be
\begin{equation}
    c_{\rm s} =  c_{\rm s0}\bigg( \frac{r_2}{M_1 + M_2}\bigg)^q \bigg( \frac{M_1}{R_1} + \frac{M_2}{R_2}\bigg)^q,
    \label{eq::EOS}
\end{equation}
where $R_1$ and $R_2$ are the radial distances from the primary and secondary stars, respectively, and
 $c_{\rm s0}$ is a constant with dimensions of velocity and $q$ is set to 3/4. This sound speed prescription guarantees that the primary and secondary stars set the temperature profiles in the circumprimary and circumsecondary discs, respectively. For $R_1, R_2 \gg r_2$, $c_{\rm s}$ is set by the distance from the centre of mass of the system.

\begin{table}
	\centering
	\caption{The setup of the SPH simulations with an initial disc around the perturber. The simulation ID is given in the first column.  The remaining columns list the tilt of the flyby orbit, $i_2$, initial tilt of the perturber disc, $i_{\rm disc,0}$, the initial mass of the perturber disc, $m_{\rm disc,0}$, and the final tilt of the perturber disc $i_{\rm disc,2}$,  along with the standard error.}
	\label{table::two_discs}
	\begin{tabular}{ccccc} 
		\hline
		Simulation & $i_2/^\circ$ & $i_{\rm disc,0}/^\circ$  & $m_{\rm disc,0}/M_{\odot}$   & $i_{\rm disc,2}/^\circ$ \\
		\hline
		PD0\_0  & $0$ & $0$ & $0.001$ & $\sim 0 \pm  0.005$ \\
		PD45\_0  & $45$ & $0$ & $0.001$ & $\sim 10 \pm 0.078$ \\
            PD45\_0\_light  & $45$ & $0$ & $0.0001$ & $\sim 30 \pm  0.072$ \\
		PD45\_45  & $45$ & $45$ & $0.001$ & $\sim 47 \pm 0.012$ \\
	    \hline
	\end{tabular}
\end{table}

\subsubsection{Perturber setup}
We vary the mass, periastron distance, tilt, and position angle of the perturber.  The standard perturber mass we select is $M_2= 0.2\, \rm M_{\odot}$, however we also use $M_2= 1\, \rm M_{\odot}$. The total mass of the system is then $M_{\rm tot} = M_1 + M_2$. The standard periastron distance we select is $r_{\rm p} = 100\, \rm au$, in which case the flyby is a grazing encounter. Simulations with the standard periastron distance will have the same periastron distance regardless of trajectory misalignment, which occurs at $x=0$ and $y>0$.  We also consider periastron distances of $r_{\rm p} = 80,\, 120,\, \rm au$. The tilt of the flyby orbit is measured with respect to the $z$--axis.  The majority of the simulations model an inclined perturber trajectory being $i_2 = 45^\circ$, but we also consider flyby orbits inclined by $0^\circ$ (coplanar prograde), $15^\circ$, $30^\circ$, $60^\circ$. A coplanar perturber initially lies in the $x$--$y$ plane and arrives from the negative $y$ direction, and leaves towards the negative $y$ direction. The reference frame within our simulations is centered on the system's center of mass.  We also model a bound companion with an eccentricities $e_2=0.3$ and $e_2=0.7$ with mass $M_2 = 0.2\, \rm M_{\odot}$. The bound companion initial begins at apastron. We simulate only a single orbital period which mimics a "flyby" encounter. The summary of the hydrodynamical simulations are given in Table~\ref{table::two_discs}.

\begin{figure} 
\centering
\includegraphics[width=1\columnwidth]{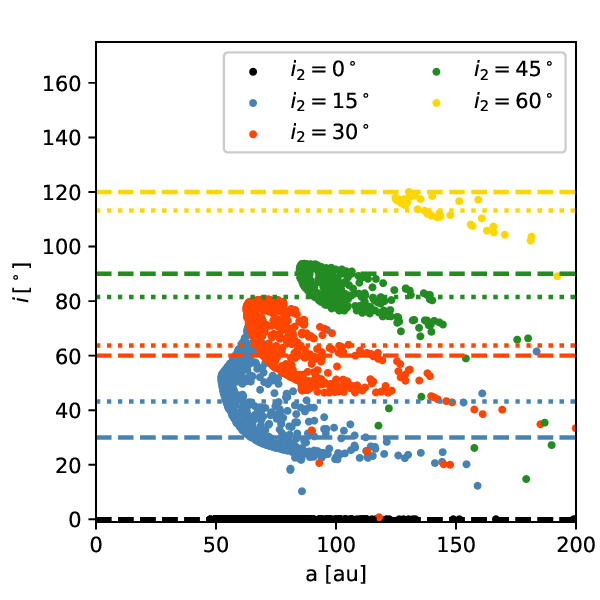}
\centering
\caption{The inclination distribution of particles captured by a parabolic flyby during our $N$--body simulations. We compare the distributions for different initial tilt of the flyby: $i_2 = 0^\circ$ (black, model N0), $15^\circ$ (blue, N15), $30^\circ$ (red, N30), $45^\circ$ (green, N45), and $60^\circ$ (yellow, N60). The horizontal-dotted lines represent the average inclination value for each model. The horizontal-dashed lines denote twice the initial perturber tilt. The particles from each simulation tend to be captured with an inclination roughly twice that of the initial perturber's tilt.}
\label{fig::nbody}
\end{figure}

\begin{figure} 
\centering
\includegraphics[width=0.88\columnwidth]{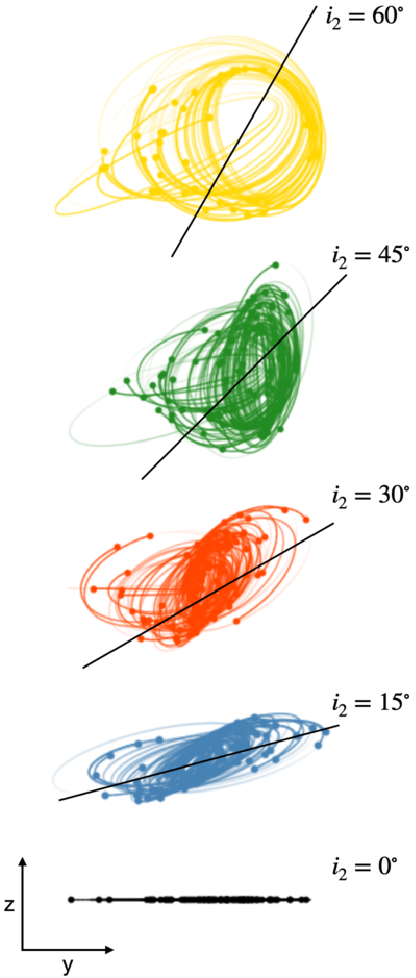}
\centering
\caption{The orbits of captured particles around flybys with different initial tilts: $i_2 = 0^\circ$ (black, model N0), $15^\circ$ (blue, N15), $30^\circ$ (red, N30), $45^\circ$ (green, N45), and $60^\circ$ (yellow, N60). The tilt of the flyby is represented by the solid black line. The solid dots denote the current position of the particles. We view the orbits in the $y$--$z$ plane. As the tilt of the perturber increases, the orbits of the majority of captured particles will have an inclination close to twice the initial perturber tilt. }
\label{fig::nbody_orbit}
\end{figure}

\subsubsection{Perturber disc setup}
In four simulations, PD0\_0, PD45\_0, PD45\_0\_light, and PD45\_45, we include an initial  circumsecondary disc around the perturber. These types of simulations aim to examine the transfer of material between two protoplanetary discs during a flyby encounter. For these simulations, we set $r_{\rm p} = 100\, \rm au$ and $M_2 = 0.2\, \rm M_{\odot}$. The circumsecondary disc mirrors the disc parameters of the primary disc (given in Section~\ref{sec::primarydisc_setup}), however, the inner and outer disc radii are set to $r_{\rm in,2} = 3.3\, \rm au$ and  $r_{\rm out,2} = 33\, \rm au$, respectively. The inner radius is chosen to equal the accretion radius of the perturber. The outer radius is chosen based on the truncation radius of a binary system being about one-third of the separation \cite[e.g.,][]{Artymowicz1994,Pichardo2005,JangCondell2015}. For PD0\_0, PD45\_0, and PD45\_45 the disc mass is set to equal the primary disc mass, $0.001\,\rm M_{\odot}$. For PD45\_0\_light, we decrease the perturber disc mass by a factor of 10, such that $m_{\rm disc,0}=10^{-4}\, \rm M_{\odot}$.  We consider two flyby orbits tilted by $0^\circ$ and $45^\circ$. For the $45^\circ$-inclined orbit, we consider three simulations, PD45\_0, PD45\_0\_light, and PD45\_45, where the disc is tilted by $0^\circ$ (misaligned to the flyby orbit) and $45^\circ$ (coplanar to the flyby orbit), respectively. For PD0\_0, PD45\_0, and PD45\_45, the initial number of SPH particles is set to $10^6$, with $500,000$ particles within the primary disc and $500,000$ particles within the perturber disc. The primary and secondary discs have a shell-averaged smoothing length per scale height of $\langle h \rangle/H \approx 0.5$. For PD45\_0\_light, the initial number of SPH particles is set to $10^6$, with $90000$ particles within the primary disc and $100,000$ particles within the perturber disc. The primary has a shell-averaged smoothing length per scale height of $\langle h \rangle/H \approx 0.4$, while the secondary disc has $\langle h \rangle/H \approx 0.8$.  The summary of the perturber disc simulations are given in Table~\ref{table::two_discs}.

\subsubsection{Analysis routine}
To analyse the hydrodynamical simulations, we average over all particles bound to either the central star or the flyby. For a particle to be bound to a particular sink, the specific energies (kinetic plus potential) of the particles are negative, neglecting the thermal energy. For each disc, we calculate the mean properties of the particles, such as the surface density, inclination (tilt), longitude of ascending node (twist), eccentricity, and mass. Similar to the $N$--body simulations, the tilt is measured with respect to the $z$--axis. We set the time $t = 0$ to represent the time of periastron passage; therefore, the initial time of the simulations will be negative.

\subsection{Limitations}
 In the context of protoplanetary discs, $N$-body simulations primarily focus on the gravitational interactions between massive bodies and test particles. However, they do not encompass the additional physics that take place within the disc. For instance, N-body simulations are indicative of a collisionless system devoid of pressure/temperature gradients and viscosity -- elements that are inherent in protoplanetary discs. While $N$-body simulations offer valuable insights into the gravitational interactions and overall dynamics of protoplanetary discs, they should be complemented with more intricate models, such as hydrodynamical simulations, which incorporate these supplementary physics (i.e., pressure, temperature gradients, and viscosity). Even a pressure-less fluid would still behave inherently different to $N$--body dynamics due to the density/velocity fields being multi-valued. Hydrodynamical simulations are vital for obtaining a comprehensive understanding of the intricate processes that mold protoplanetary discs during flyby encounters. It is noteworthy, however, that our hydrodynamical simulations still possess certain limitations, notably pertaining to resolution, a topic discussed in Appendix~\ref{appendix::resolution}. Furthermore, in hydro-simulations of stellar flybys with discs it is also important to account for radiation effects – especially for disc-penetrating encounters \cite[as in live radiative calculations by][]{Borchert2022a,Borchert2022b}. These effects are expected to modify the stellar accretion, the 3D-temperature field, and the disc aspect ratio during the encounter. However, the orbital plane of the captured material for disc-grazing encounters (relevant for this work) is expected to remain unchanged. The presence of gas unavoidably damps orbital oscillations that would survive in pure $N$--body simulations.




\begin{figure*}
\includegraphics[width=2\columnwidth]{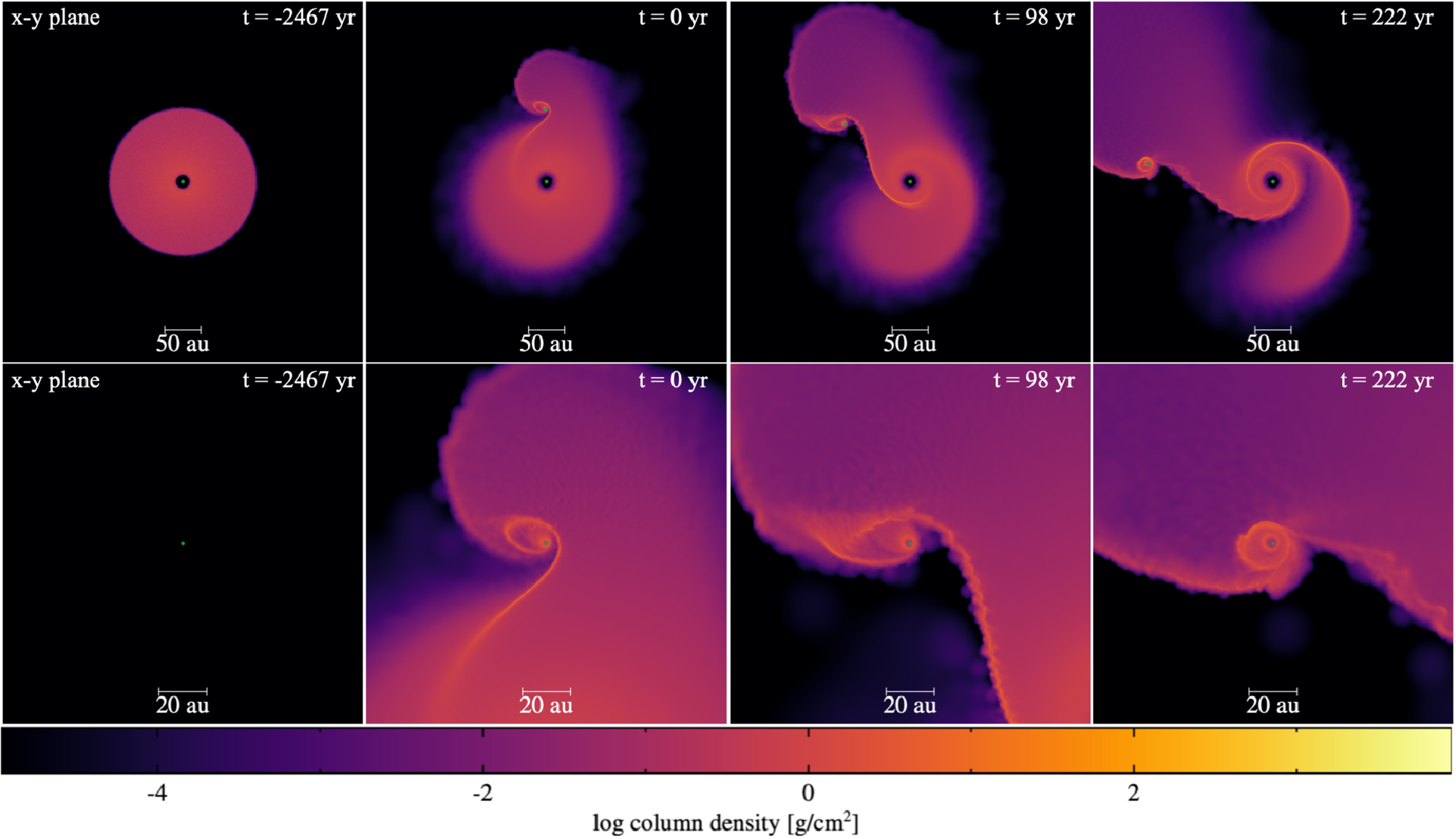}\centering
\caption{The evolution and formation of protoplanetary discs around the primary star and perturber (green dots) during a coplanar prograde encounter (model H0). The frames in the top row are centered on the primary star, while the frames in the bottom row are zoomed-in and centered on the flyby. All the frames are viewed in the  $x$--$y$ plane, which is face-on to the primary disc.  The first column shows the primary disc and perturber at the beginning of the simulation. The second column shows the disc structure during the periastron passage of the flyby ($t = 0\, \rm yr$). The third and fourth columns represent times shortly after the periastron passage, indicating the formation of the disc around the flyby. The color denotes the disc surface density. 
}
\label{fig::coplanar_splash}
\end{figure*}

\begin{figure} 
\centering
\includegraphics[width=0.98\columnwidth]{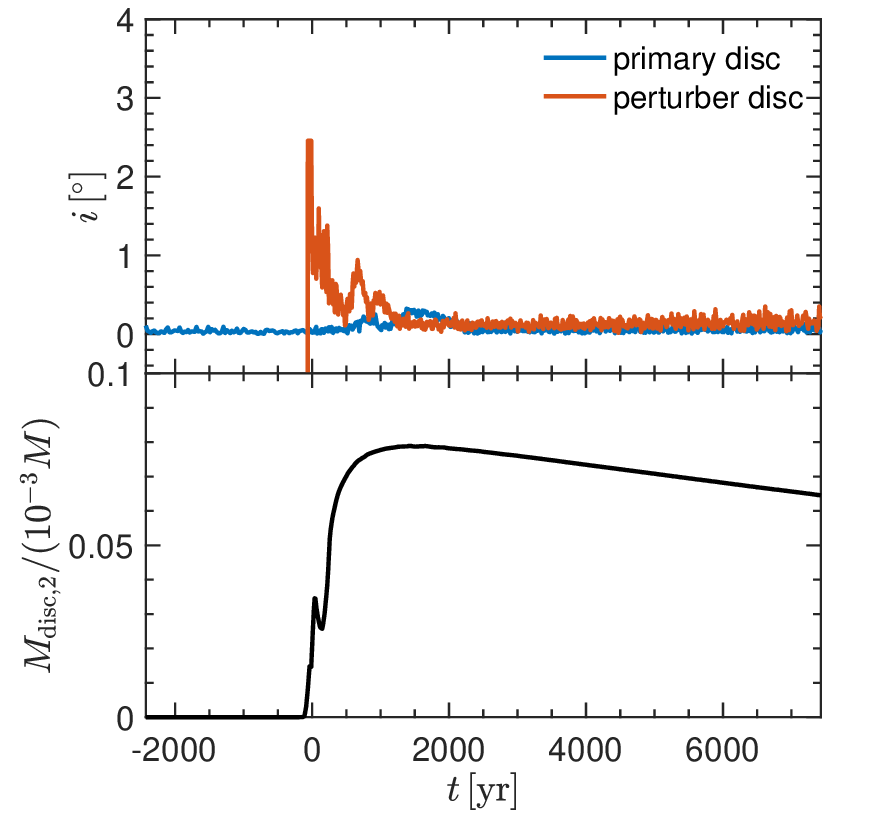}
\centering
\caption{Top panel: the tilt of the primary disc (blue) and perturber disc (red) for a coplanar perturber (model H0). Bottom panel: the fraction of the perturber disc mass to the initial primary disc mass. The captured material around the perturber is nearly in a coplanar orientation.}
\label{fig::coplanar_tilt_mass}
\end{figure}

\section{\texorpdfstring{$N$}{}--body results}
\label{sec::Nbody_results}
Here, we analyse the mass transfer during a parabolic encounter using $N$--body numerical simulations. Previous works have simulated the interaction  between a particle disc and a flyby with $N$--body calculations \cite[e.g.,][]{Clarke1993,Hall1996,Larwood2001,Pfalzner2005a,Jikova2016}. In particular, \cite{Jikova2016} found that the perturber tilt affected the captured particles' tilt distribution. However, they did not detail the relationship between the captured particles' tilt and the flyby's initial tilt.  We further analyse this by conducting $N$--body simulations with various initial tilts of the perturber using {\sc rebound}. 

Figure~\ref{fig::nbody} shows the inclination distribution of particles captured by the flyby. We simulate different initial tilts of the perturber, $i_{2} = 0^\circ$, $15^\circ$, $30^\circ$, $45^\circ$, and $60^\circ$. For the coplanar encounter, $i_{2} = 0^\circ$, the particles are captured with a coplanar tilt. For each inclined case, the resulting captured particles have an inclination distribution approximately twice the initial perturber tilt. To clarify this, we plot a horizontal  dashed line at twice the initial perturber tilt for each case. For example, for $i_2 = 45^\circ$, the captured particles have tilts that are $\sim 90^\circ$ with respect to the tilt of the primary disc. For $i_2 = 60^\circ$, fewer particles are captured, but the captured particles have tilts that are $\sim 120^\circ$, which are considered retrograde orbits. In general, we find as the tilt of the perturber increases, fewer particles are captured. Thus, lower inclination encounters are more efficient at capturing material.  
 This is consistent with the results presented in \cite{Jikova2016}. Figure~\ref{fig::nbody_orbit} shows the orbits of the material captured around the flyby after periastron passage. The inclinations shown in this plot are the same as the ones shown in Fig.~\ref{fig::nbody}.

\begin{figure*} 
\includegraphics[width=2\columnwidth]{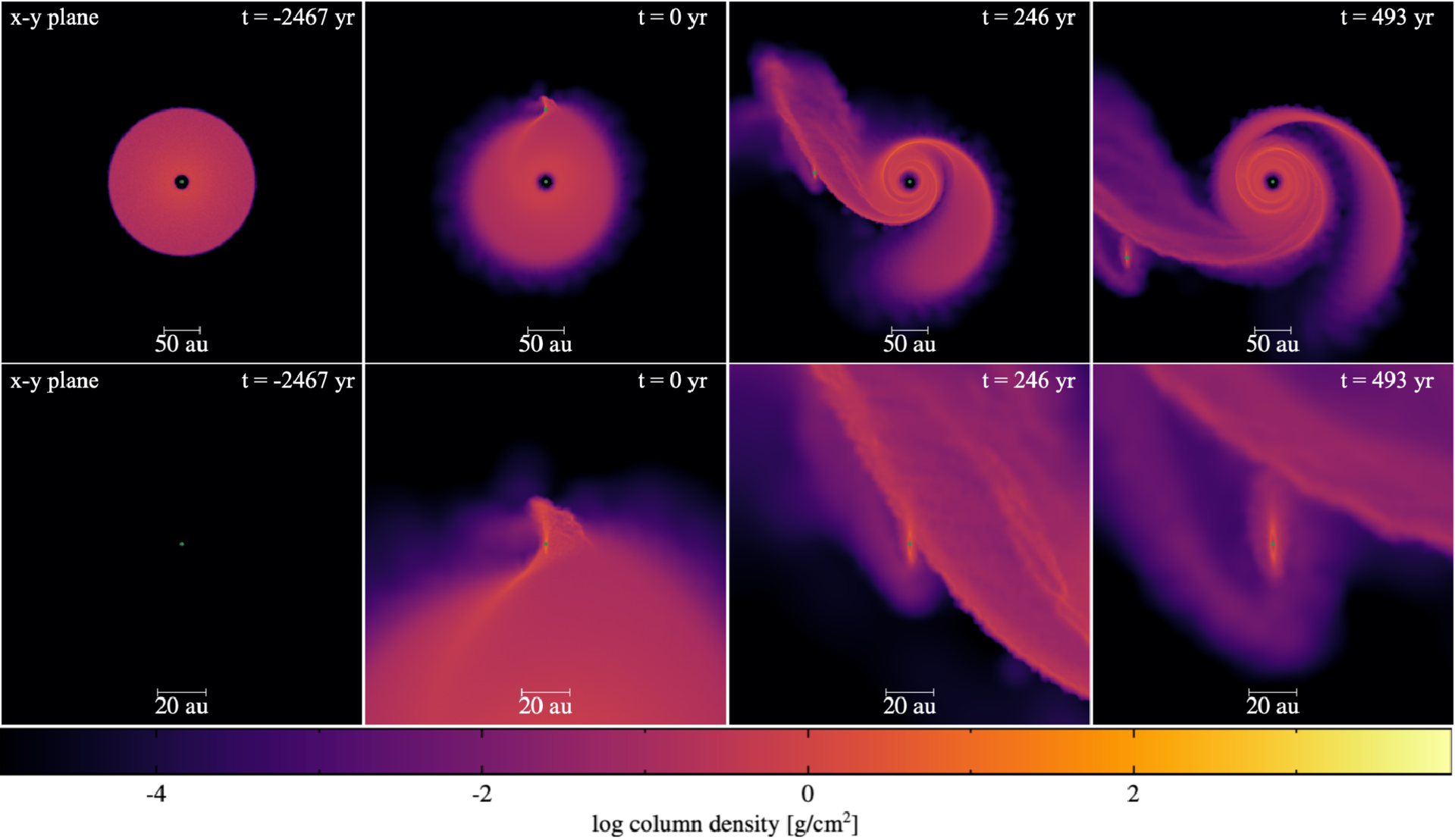}\centering
\caption{Same as Fig.~\ref{fig::coplanar_splash} but for a $45^\circ$ inclined perturber (model H45).  }
\label{fig::inclined_splash}
\end{figure*}

\begin{figure} 
\centering
\includegraphics[width=\columnwidth]{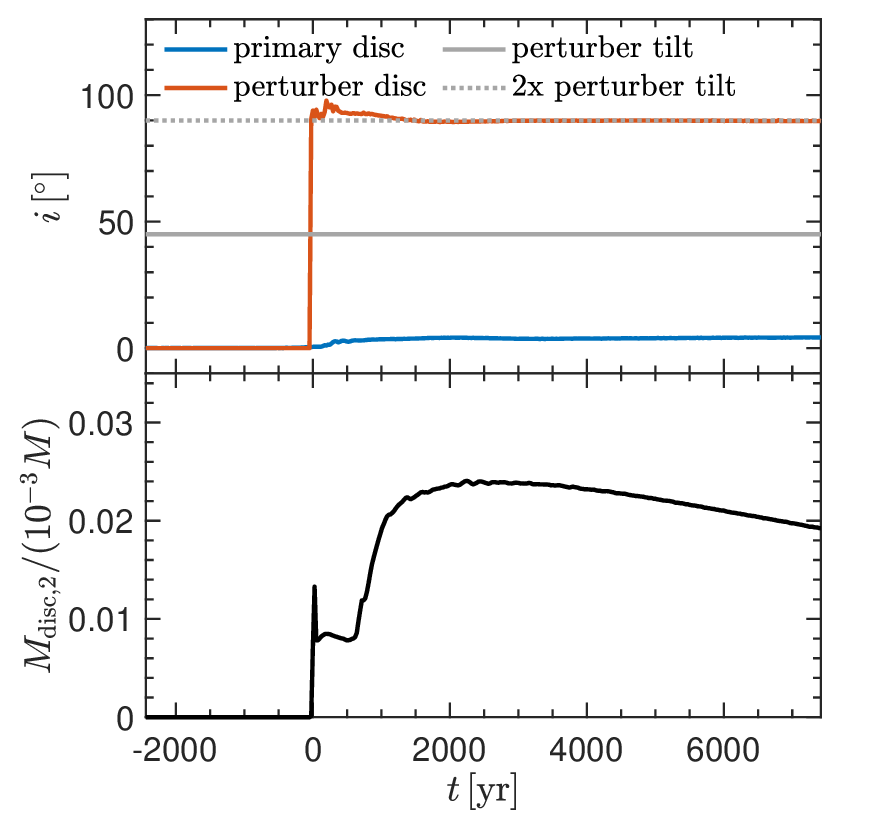}
\centering
\caption{Same as Fig.~\ref{fig::coplanar_tilt_mass} but for a $45^\circ$ inclined perturber (model H45). The tilt of the perturber is shown by the horizontal gray line, and  twice the perturber tilt is shown by the horizontal dotted gray line. 
}
\label{fig::45_tilt_mass}
\end{figure}

\section{Hydrodynamical results}
\label{sec::hydro_results}

\subsection{Coplanar prograde flyby}
We first consider a flyby on a coplanar parabolic orbit (model H0 from Table~\ref{table::setup}).  Figure~\ref{fig::coplanar_splash} shows the evolution of this simulation, where the top row shows the interaction between the coplanar perturber and the primary disc, and the bottom row shows a zoomed-in view centered on the perturber. The second column displays the disc structure when the perturber is at the periastron. At this point, the perturber captures material from the primary disc as gaseous streams. The streams flow around the perturber, forming a disc (seen clearly in the zoomed-in panel). The third and fourth columns display the structure of the two protoplanetary discs shortly after the periastron passage. At these times, a gaseous stream still supports the growth of the forming disc around the perturber.

We now investigate the structure of the perturber disc in more detail. The upper panel in Fig.~\ref{fig::coplanar_tilt_mass} shows the tilt evolution for the primary and perturber discs. The primary disc tilt is initially coplanar and maintains a coplanar profile during and after the encounter. During periastron passage, a disc forms around the perturber that initially forms at a tilt of $\sim 5^\circ$, but then quickly damps to a coplanar orientation, consistent with the $N$--body simulations. The bottom panel in   Fig.~\ref{fig::coplanar_tilt_mass} shows the ratio of the perturber disc mass to the initial primary disc mass. Shortly after periastron passage, the perturber disc is at peak mass, which is about $10$ per cent of the primary disc mass. The secondary disc's mass decreases over time from material accreting onto the perturber.

\subsection{\texorpdfstring{$45^\circ$}{}--inclined flyby}
\label{sec::inclined_flyby}

 In this section, we progress from a simplified coplanar encounter to a more probable inclined encounter. Perfectly coplanar/aligned flybys are less likely than inclined ones, which can be either prograde or retrograde. Figure~\ref{fig::inclined_splash} shows the evolution of simulation H45 (a flyby tilted by $45^\circ$). The perturber captures material that forms a disc. However, in this case the disc appears perpendicular to the primary disc in the $x-y$ plane. A resolution study for this specific simulation is given in Appendix~\ref{appendix::resolution}.
 


The upper panel in Fig.~\ref{fig::45_tilt_mass} shows the tilt evolution for the primary and perturber discs. The primary disc tilt is initially coplanar but increases to $\sim 4^\circ$ as a consequence of the flyby encounter. The primary disc maintains this increased tilt for the duration of the simulation. The periastron passage of the flyby occurs at $\sim 2400\,\rm yr$.  At this time, a disc forms around the perturber with an initial tilt of $\sim 98^\circ$, but damps to $\sim 90^\circ$, which is twice the tilt of the perturber (given by the dotted-horizontal line). Therefore, the secondary disc does not form at the same tilt as the perturber orbit but forms a factor of two larger.  This is consistent with our $N$--body simulations (see Section~\ref{sec::Nbody_results}). Moreover, the mutual inclination between the primary and secondary discs is $\sim 90^\circ$. The bottom panel in  Fig.~\ref{fig::45_tilt_mass} shows the ratio of the perturber disc mass to the primary disc mass. Shortly after periastron passage, the perturber disc grows to peak mass, which is about $3$ per cent of the primary disc mass, then decreases over time from material accreting onto the perturber.

\begin{figure} 
\centering
\includegraphics[width=\columnwidth]{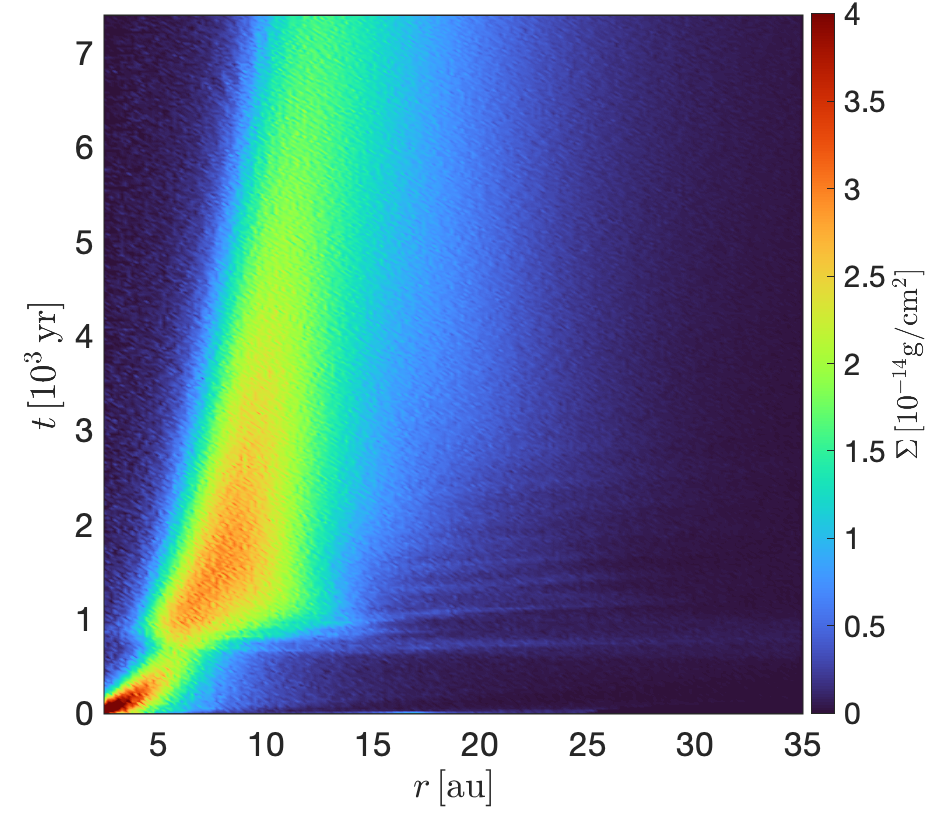}
\centering
\caption{ The surface density evolution for the second-generation disc. The $x$ shows the disc radius, while the $y$--axis shows the time with $t = 0 \, \rm yr$ being the time of periastron passage. The colour denotes the surface density.}
\label{fig::sigma_r}
\end{figure}

\begin{figure} 
\centering
\includegraphics[width=\columnwidth]{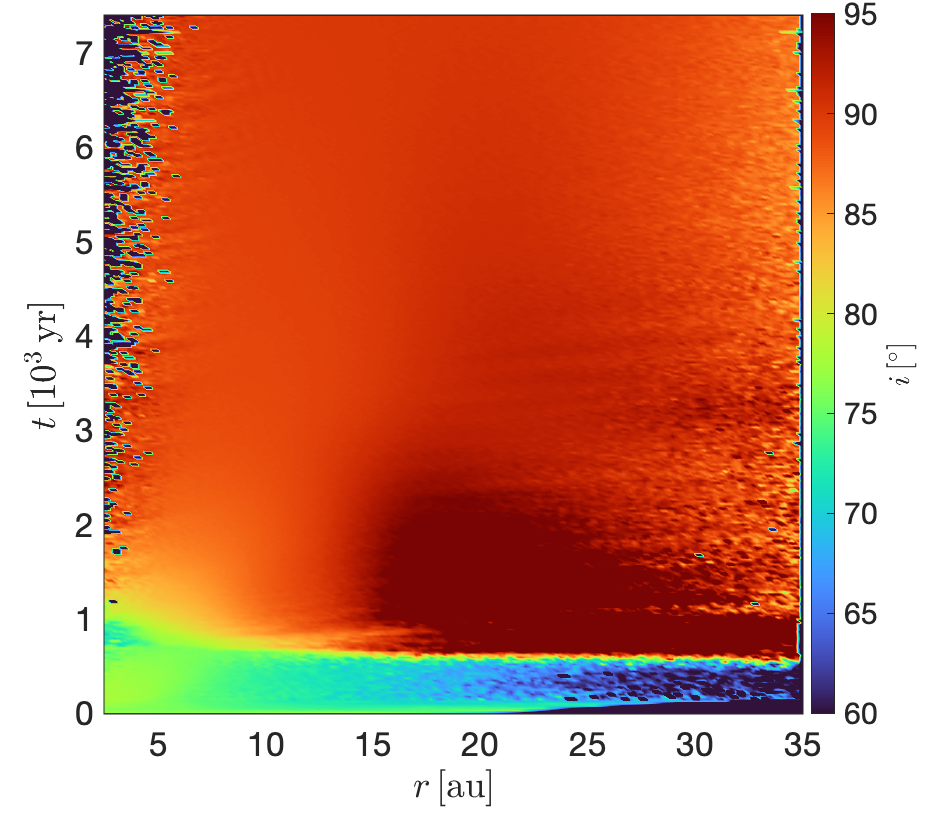}
\centering
\caption{ The tilt evolution for the second-generation disc. The $x$ shows the disc radius, while the $y$--axis shows the time with $t = 0 \, \rm yr$ being the time of periastron passage. The colour denotes the tilt. }
\label{fig::tilt_r}
\end{figure}

\begin{figure} 
\centering
\includegraphics[width=\columnwidth]{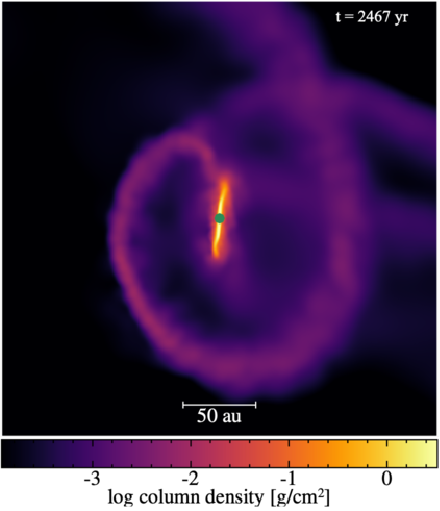}
\centering
\caption{The formation of second-generation disc around the perturber (green dot) during a $45^\circ$-inclined flyby (model H45). Multiple gaseous streamers are present with the more prominent streamer accreting material at a higher inclination than the less prominent streamers.}
\label{fig::streams}
\end{figure}

\begin{figure*} 
\centering
\includegraphics[width=\columnwidth]{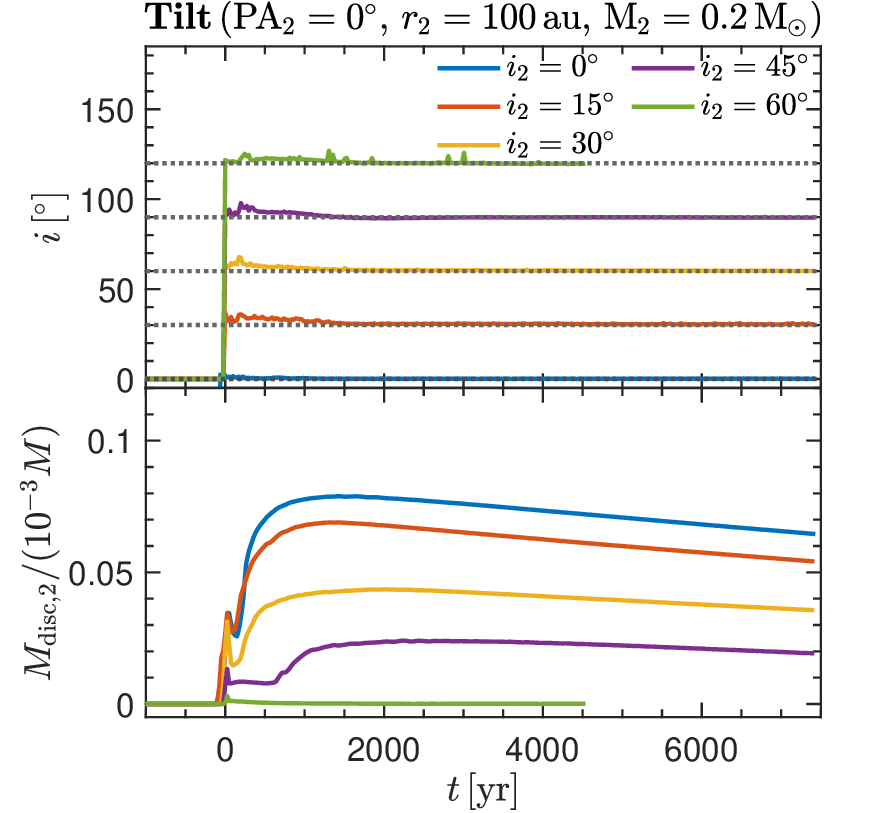}
\includegraphics[width=\columnwidth]{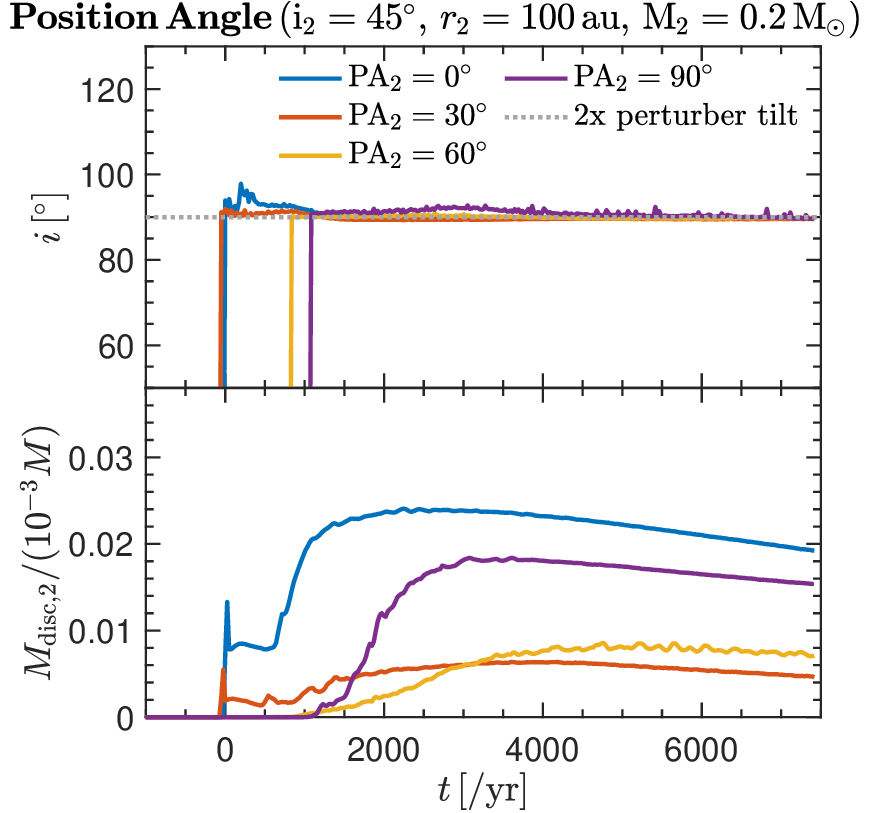}
\includegraphics[width=\columnwidth]{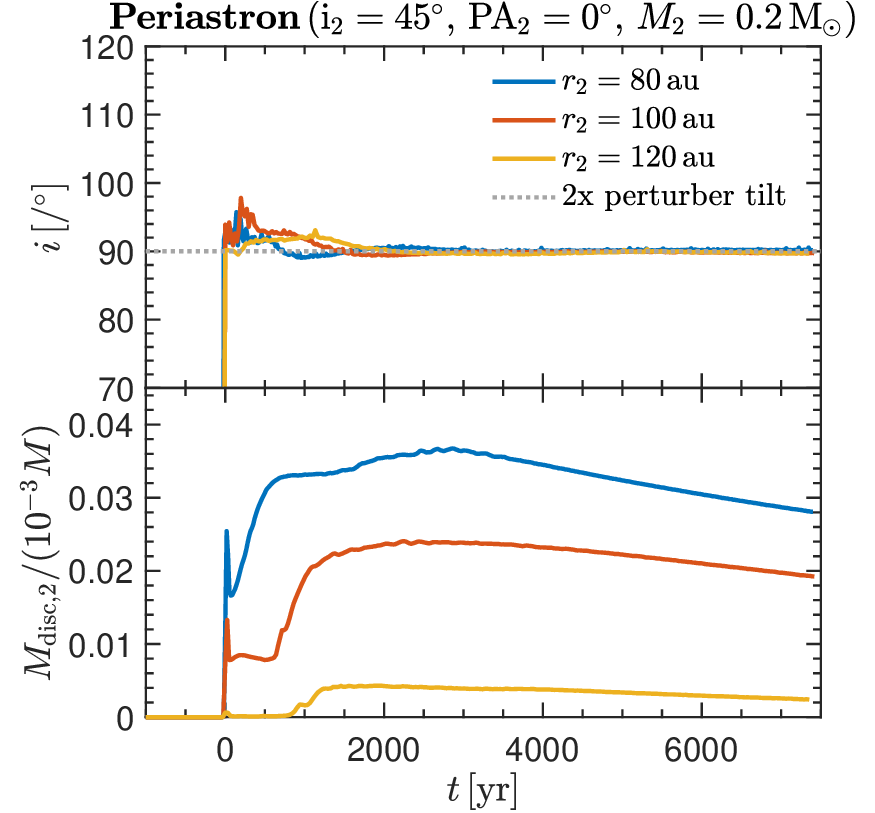}
\includegraphics[width=\columnwidth]{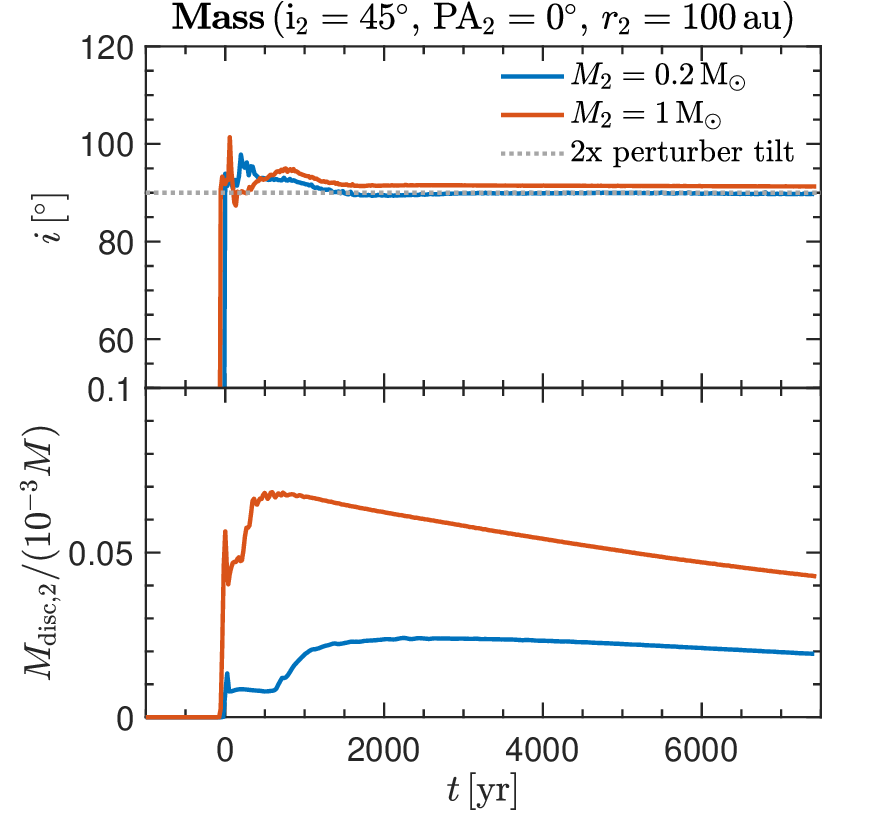}
\centering
\caption{The evolution of the tilt and mass of the disc forming around the perturber for different parameters of the perturber: tilt (top-left panel), position angle (top-right panel), periastron distance (bottom-left panel), and mass (bottom-right panel).  For the position angle, periastron distance, and mass, the flyby orbit is set to $45^\circ$. The horizontal lines show twice the initial tilt of the respected flyby orbits. The analysis of the disc around the $60^\circ$-inclined perturber due to low disc resolution from the lower amount of captured material. In each case, the forming disc around the flyby is captured at twice the initial perturber tilt.}
\label{fig::flyby_data}
\end{figure*}

 Figure~\ref{fig::sigma_r} shows the surface density evolution of the disc around the flyby. At the end of the simulation, the spatial size of the disc extends from  $\sim 5\, \rm au $  to $\sim 30\, \rm au$, with the peak of the surface density profile located at $\sim 12\, \rm au$. The surface density profile goes as $\Sigma \propto r^{-3/2}$. From Fig.~\ref{fig::45_tilt_mass}, we measure the density-weighted average of the disc tilt. We check to see whether the average disc tilt calculated encompasses the entire spatial size of the disc. Figure~\ref{fig::tilt_r} shows the tilt evolution as a function of disc radius ($x$--axis) and time ($y$--axis). At $t \lesssim 1000\, \rm yr$, the tilt of the disc is dominated by material accreting onto the disc at a lower tilt, while at $ 1000\, \rm yr\lesssim t \lesssim  2500\, \rm yr$ the tilt is dominated by material accreting onto the disc at a higher tilt. Beyond $\sim 3000\, \rm yr$ after the periastron passage, the disc has a tilt of twice the initial perturber tilt at all radii. In Fig.~\ref{fig::streams}, we take a closer look at the infall onto the circumsecondary disc around the flyby. There are three streams of material accreting onto the disc. At this time, the more predominant streamer is accreting material at a higher inclination than the other two less predominant streamers.

\subsection{Varying flyby parameters}
In this subsection, we vary the tilt, position angle, periastron distance, and mass of the flyby to explore the robustness of the flyby disc forming at a tilt twice the initial flyby tilt, in the latter three experiments we keep the tilt at $45^\circ$. Our primary focus centers on quantifying the tilt of the disc. However, to provide a comprehensive analysis, we have also included an examination of the resulting disc phase angle in Appendix~\ref{appendix::phase_angle} for all simulations.

\begin{figure*} 
\includegraphics[width=2\columnwidth]{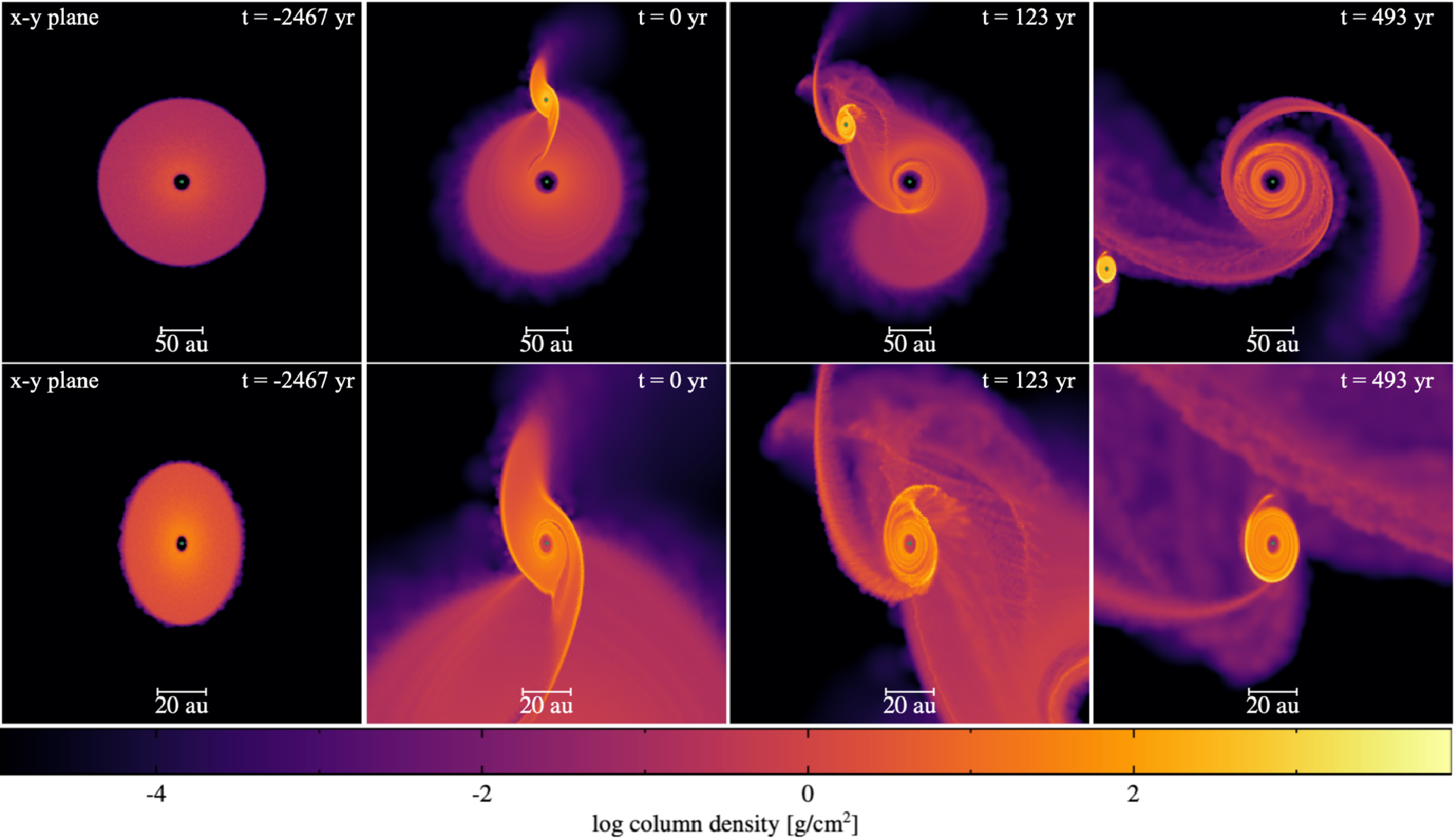}\centering
\caption{The evolution of two interacting protoplanetary discs around the primary star and perturber (green dots) during a coplanar prograde encounter (model PD45\_45 ). The frames in the top row are centered on the primary star, while the frames in the bottom row are zoomed-in and centered on the flyby. All the frames are viewed in the  $x$--$y$ plane, which is face-on to the primary disc.  The first column shows the primary disc and perturber at $t=0\, \rm yr$. The second column shows the disc structure during the periastron passage of the flyby. The third and fourth columns represent times shortly after the periastron passage, indicating the formation of the disc around the flyby. The color denotes the disc surface density. }
\label{fig::fd3_splash}
\end{figure*}

\subsubsection{Flyby tilt}
We analyze how the initial flyby tilt affects the tilt of the forming disc around the flyby. The initial flyby tilts are $i_0 = 0^\circ$, $15^\circ$, $30^\circ$, $45^\circ$, and $60^\circ$.  The top-left panel in Fig.~\ref{fig::flyby_data} shows the tilt and mass of the disc around the flyby as a function of time. The horizontal dotted lines represent twice the initial perturber tilt for initial tilts $i_2 = 0^\circ$ to $60^\circ$. When the perturber orbital tilt is $60^\circ$, the disc forms retrograde at $\sim 120\rn{^\circ}$.  For this model, we only analyze the disc up to $7000\, \rm yr$ due to low disc resolution because of less material captured by the perturber.
The mass of the perturber disc decreases with increasing flyby tilt, with the coplanar flyby resulting in the highest disc mass. There is also a delay in the time of peak perturber disc mass and the time of periastron passage, which is shorter as the tilt of the perturber decreases. By varying the flyby tilt, the forming disc around the perturber still forms at a tilt twice the perturber tilt.

\subsubsection{Flyby position angle}
Next, we vary the position angle of the flyby orbit. We consider position values $\rm PA_2 = 0^\circ$, $30^\circ$, $60^\circ$, and $90^\circ$. The top-right panel in Fig.~\ref{fig::flyby_data} shows the perturber disc tilt and mass as a function of time for the different position angle models. The horizontal dotted line represents twice the initial perturber tilt of $45^\circ$. The perturber disc is captured at a tilt of $\sim 90^\circ$ regardless of the position angle of the flyby. A $\rm PA_2 = 0^\circ$ flyby results in the highest disc mass out of all the PA simulations. When $\rm PA_2 = 30^\circ$ and $60^\circ$, the disc mass is similar with a mass of $\sim 1$ per cent of the primary disc mass. For $\rm PA_2 = 90^\circ$, the disc mass is $\sim 2$ per cent of the primary disc mass. This is because when $\rm PA_2 = 90^\circ$, the flyby has two closest approaches on either side of the primary disc, capturing more material. By varying the flyby position angle, the forming disc around the perturber still forms at a tilt twice the perturber tilt.

\subsubsection{Flyby periastron}
Next, we vary the periastron distance of the flyby orbit. We consider position values $r_{\rm p} = 80\, \rm au$, $100\, \rm au$, and $120\, \rm au$.   The bottom-left panel in Fig.~\ref{fig::flyby_data} shows the perturber disc tilt and mass as a function of time for the different periastron distance models. The horizontal dotted line represents twice the initial perturber tilt of $45^\circ$. For periastron distances, $r_{\rm p} = 80\, \rm au$, $100\, \rm au$, and $120\, \rm au$, the tilt of the perturber disc is $90^\circ$ (twice the initial perturber tilt) with respect to the $z$--axis. For $r_{\rm p} = 80\, \rm au$, the perturber penetrates the disc, resulting in the highest disc mass compared to the periastron distance simulations. Moreover, as the periastron distance of the perturber increases, the resulting disc mass decreases. By varying the flyby periastron distance, whether a grazing or lightly penetrating encounter, the forming disc around the perturber still forms at a tilt twice the perturber tilt.

\subsubsection{Flyby mass}
Finally, we vary the mass of the flyby. We consider mass values of $M_2 =  0.2\, \rm M_{\odot}$ and $1\, \rm M_{\odot}$.  The bottom-right panel in Fig.~\ref{fig::flyby_data} shows the perturber disc tilt and mass as a function of time for the different flyby mass models. The horizontal dotted line represents twice the initial perturber tilt of $45^\circ$.  For $M_2 = 0.2\, \rm M_{\odot}$ and $1\, \rm M_{\odot}$, the disc forming around the perturber has  a tilt slightly larger than twice the perturber tilt. The more massive perturber captures more material, resulting in a higher disc mass of $\sim 10$ per cent of the primary disc mass.  We can see that varying the flyby mass does have a small affect on the final disc inclination, but the disc that forms still has roughly twice the initial perturber tilt.


\section{Interacting protoplanetary discs}
\label{sec::two_discs}
This section explores situations where the perturber initially has a protoplanetary disc before interacting with the primary disc. We simulate three combinations of the perturber and disc around the perturber (given in Table~\ref{table::two_discs}) which are 1) a coplanar flyby with a disc coplanar to the flyby orbit, 2) a $45^\circ$-inclined flyby with a disc coplanar to the primary disc, and 3) a $45^\circ$-inclined flyby, with a disc coplanar to the flyby orbit.  Figure~\ref{fig::fd3_splash} shows the disc surface density for the two interacting protoplanetary discs for a $45^\circ$-inclined flyby, with a $45^\circ$-tilted disc (model PD45\_45). The top row is centered on the primary disc, and the bottom row is centered on the disc around the perturber. The first column represents the initial structure of the two discs. The second column shows the time of periastron passage of the perturber. The two discs interact with one another, where material from the secondary disc is transferred to the primary and vice versa. The third and fourth columns show times shortly after the periastron passage. Gaseous streams from the primary disc are accreting onto the perturber disc. 

\begin{figure} 
\includegraphics[width=1\columnwidth]{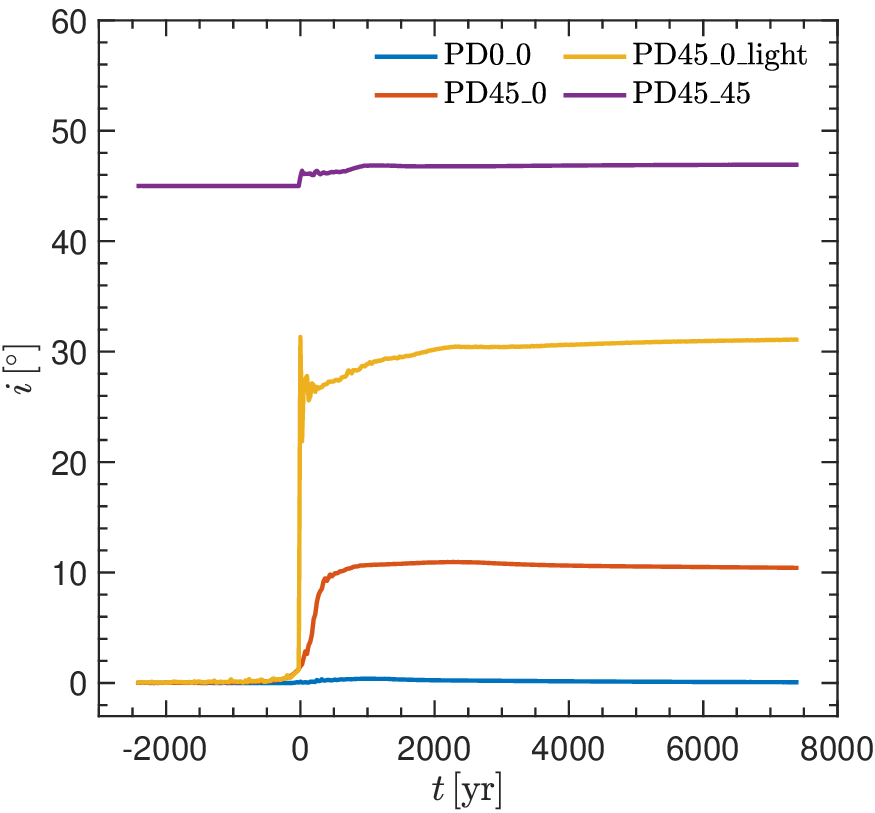}\centering
\caption{A summary of the perturber disc tilt as a function of time for the four interacting protoplanetary disc simulations. We show the models PD0\_0 (blue), PD45\_0 (red), PD45\_0\_light (yellow) and PD45\_45 (purple).}
\label{fig::two_disc_tilt}
\end{figure}

Next, we look at the change in the tilt of the perturber disc after interacting with the primary disc. Figure~\ref{fig::two_disc_tilt} shows the tilt profile as a function of time for the four models of interacting protoplanetary discs, PD0\_0 (blue), PD45\_0 (red), PD45\_0\_light (yellow) and PD45\_45 (purple). During a coplanar interaction (PD0\_0), the secondary disc remains coplanar after interacting with the primary disc. For an inclined flyby with a coplanar disc (PD45\_0), the coplanar disc increases to a tilt of $\sim 10^\circ$ after interacting with the primary disc. For an inclined flyby with a coplanar low-mass disc (PD45\_0\_light) increases to a tilt of $\sim 30^\circ$ after interacting with the primary disc. Lastly, for an inclined flyby with a $45^\circ$-tilted disc (PD45\_45), the tilt of the secondary disc increases a small amount to $\sim 47^\circ$. Unlike the simulations without an initial secondary disc,  there is no straightforward relationship between the perturber orbital tilt and the secondary disc tilt. This result is strongly dependent on the balance of the initial flyby disc angular momentum to the angular momentum of captured particles. The simulations described above have an initial circumsecondary disc around the flyby with an angular momentum equal to the primary disc. If the angular momentum of the flyby disc is significantly less than the angular momentum of the captured particles, the disc around the flyby should form at a different tilt than the original tilt.



\section{Why a factor of two?}
\label{sec::analytics}
Both our N-body and SPH calculations have motivated that the captured material has an inclination twice that of the encounter. Importantly, this result appears robust to changes in the mass of the perturber, the inclination of the encounter, the distance of closest approach and the position angle of the flyby. Here we will provide an analytic framework for this behaviour that is informed by our previous simulations.

The step-function nature of the inclination in Figure~\ref{fig::45_tilt_mass} demonstrates that the inclination of the gas does not appreciably change after the interaction. That is, the angular momentum of the material that finishes around the perturber is what it has \emph{at the instant} it is captured during the pericentre passage. We can thus calculate the properties of the gas while it is in the disc and safely assume that those properties will broadly hold as the captured material is carried away by the perturber. As the relative orientation of the gas is determined by its angular momentum, we will focus on a description of this here.

First we consider the gas in the primary disc. From the primary star, the distance to a particular region of the disc is given by $\mathbfit{r}_1$, where
\begin{equation}
    \mathbfit{r}_1 = (r_1 \cos\phi, r_1 \sin\phi, 0),
\end{equation}

\begin{figure*} 
\centering
\includegraphics[width=1\textwidth]{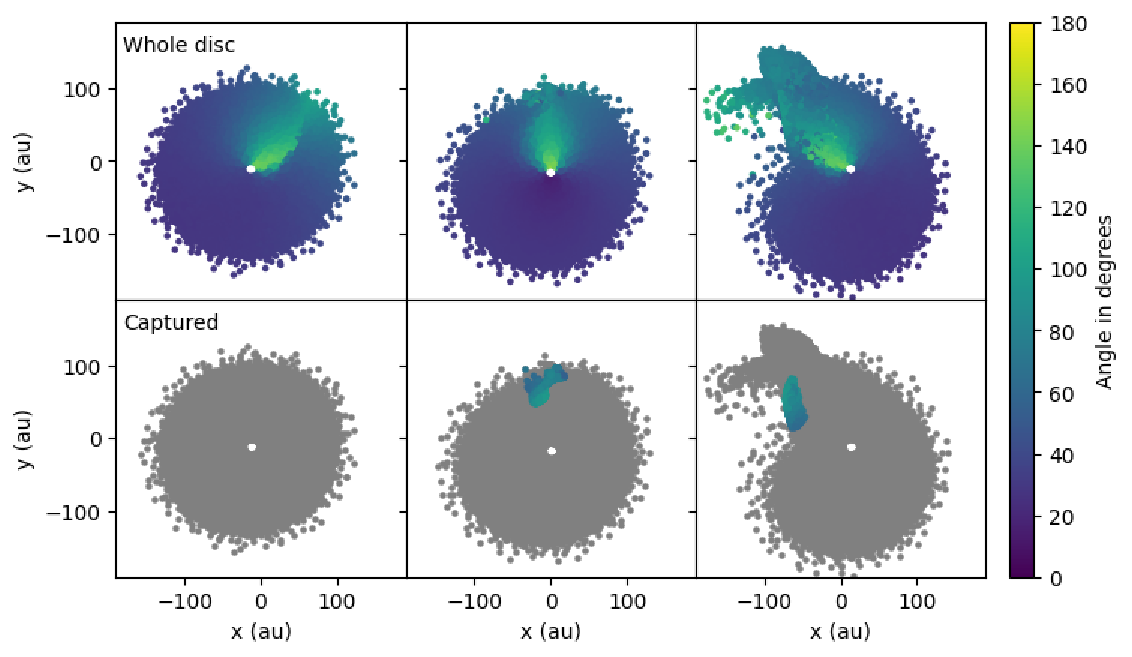}
\centering
\caption{Inclination of the gas in the disc from the frame of the perturber for simulation H45HR. \emph{Upper panel}: inclination of every particle in the disc assuming it was in orbit around the perturber. \emph{Lower panel}: only particles captured in this time-step are indicated, showing where these particles come from. The relative inclination of the gas particles varies across the disc and depends on the location of the perturber. Particles captured by the perturber are serendipitously selected from the region that always corresponds to roughly a factor of twice the original inclination.}
\label{fig::analytical_disc}
\end{figure*}
and $\phi$ is the angle measured from the point of closest approach between the perturber and primary star (see the left panel in Fig~\ref{fig::flyby_orbit}). If the disc is otherwise undisturbed the material will have a Keplerian rotation profile given by
\begin{equation}
    \mathbfit{v}_1 = (-v_{\rm{Kep}}\sin\phi,v_{\rm{Kep}}\cos\phi,0),
\end{equation}
with $v_{\rm{Kep}} = \sqrt{GM_1/r_1}$. Second, we consider the motion of the perturber. Our perturber approaches on an inclined path defined by the angle $i_{\rm 0}$, measured from the midplane of the primary disc. The path of the perturber with respect to the primary is then described as \citep[e.g.][]{DOnghia2010},
\begin{equation}
    \mathbfit{r}_2 = (-r_2 \cos i_{\rm 0} \sin i_{\rm 0}, r_2\cos i_{\rm 0}, r_2\sin i_{\rm 0} \sin i_{\rm 0}).
\end{equation}

Similarly, the perturber has a velocity given by
\begin{equation}
    \mathbfit{v}_2 = (-v_0\cos i_{\rm 0}, 0, v_0 \sin i_{\rm 0}),
\end{equation}
with $v_0 = \sqrt{2G(M_1 + M_2)/r_{\rm p}}$, where $r_{\rm p}$ is the closest approach distance. During the encounter, the perturber imparts an impulse to the gas in the disc which we name $\Delta \mathbfit{v}$. We follow the method outlined in \citet{DOnghia2010} to calculate this velocity perturbation driven by an inclined, parabolic flyby. We refer the interested reader to Appendix~\ref{appendix::analytics} for the full form of $\Delta \mathbfit{v}$.

Finally, we consider the velocity and position of the gas in the disc with respect to the perturber. Straightforwardly,
\begin{align}
    \mathbfit{R} = \mathbfit{r}_1 - \mathbfit{r}_2,
    \label{equation::r2}
\end{align}
and
\begin{align}
    \mathbfit{V} = \mathbfit{v}_1 + \Delta \mathbfit{v} - \mathbfit{v}_2.
    \label{equation::v2}
\end{align}

We find for the parameters chosen in our problem $|\Delta \mathbfit{v}| \ll |\mathbfit{V}|$ for all typical combinations of the perturber properties (inclination, mass, pericentre distance, position angle, etc.). This suggests that the velocity of the gas in the disc as measured from the perturber is effectively only dependent on perturber properties; the mass of the primary, the total mass of the stars, the inclination of the encounter and the distance of closest approach. We can thus use Equations~\ref{equation::r2} and \ref{equation::v2} to calculate the angular momentum of the gas with respect to the perturber and as a result, measure the inclination of the material with respect to the perturber.

From these two expressions we can calculate the angular momentum of the gas, $\mathbfit{L}_2 = m(\mathbfit{R} \times \mathbfit{V})$ averaged across each particle bound to the perturber, at any point during the encounter. The upper panels of Figure~\ref{fig::analytical_disc} show this for the fiducial calculation with $i_0 = 45^{\circ}$. Measured from the frame of the perturber, the inclination of the gas varies between $0^{\circ}$ and $\sim 120^{\circ}$ with higher inclinations on the side closest to the perturber. 

In the lower panels of Figure~\ref{fig::analytical_disc} we show the inclination of the gas that is captured at each snapshot. By only highlighting these particles it is clear that the inclination of the captured material \emph{at the instance of capture} is $\sim70 - 90^{\circ}$. More importantly, as the simulation evolves the region where particles are able to be captured from moves such that material with roughly the same inclination is captured at different time-steps. The serendipitous capture of particles from a region that has roughly twice the inclination of the encounter appears to be the cause of the factor of two identified across all of our simulations. 

\begin{figure} 
\includegraphics[width=1\columnwidth]{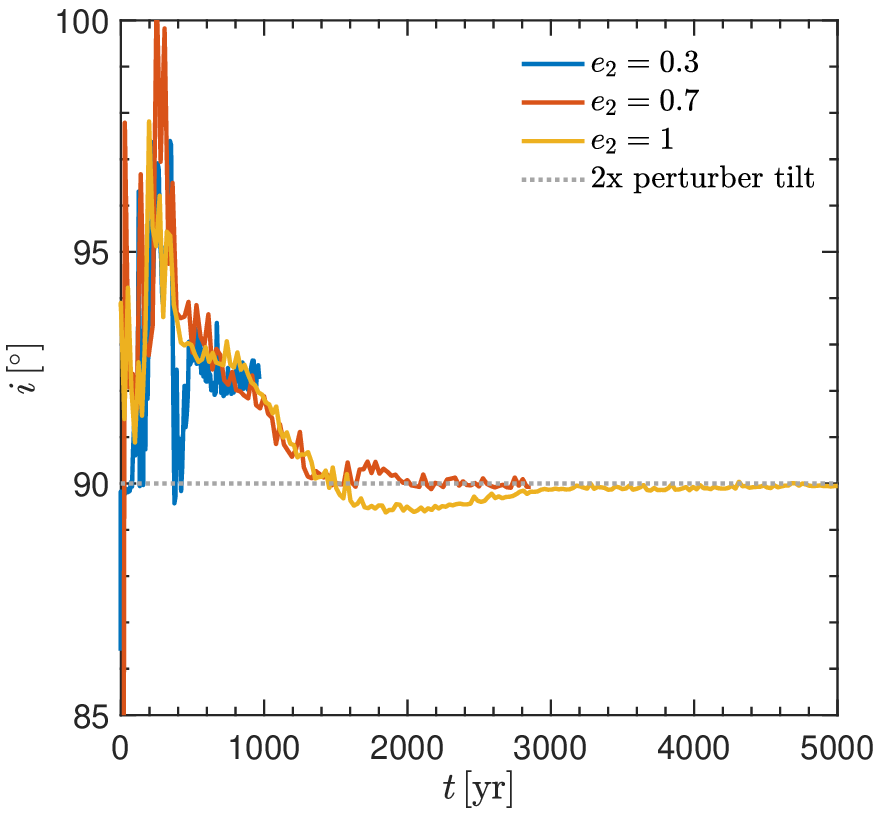}\centering
\caption{The tilt of the captured disc around the perturber with different eccentricities, $e_2 = 0.3$ (blue,  H45p3), $0.7$ (red,  H45p5), and $1$ (yellow, H45). The bound cases are only simulated for one orbit, which imitates a flyby. Therefore, the dotted lines show the data extrapolated to the simulation end-time of H45. The perturber with eccentricities $e_2=0.7$ and $1$ form a disc with a tilt twice the initial perturber tilt (gray dotted line). The perturber with a lower eccentricity, $e_2 = 0.3$, forms a disc that is not exactly twice the initial perturber tilt.}
\label{fig::ecc_tilt}
\end{figure}

The capture of this material depends on the relative velocity, so we further test this relationship by conducting additional simulations with different approach speeds. Fig.~\ref{fig::ecc_tilt} shows how the inclination of the captured disk around the perturber varies with different eccentricities: $e_2 = 0.3$ (blue, H45p3), $0.7$ (red, H45p5), and $1$ (yellow, H45). The simulations for the bound cases are conducted for a single orbit, which imitates a flyby scenario. The three curves are indistinguishable up to their individual cut off points. While the relative velocity between the primary disc and perturber is responsible for determining the factor of two in inclination, we find that even a large change in the energy of the encounter produces only a small change in the relative inclination.



\section{Discussion}
\label{sec::discussion}
 Our results demonstrate that the relative inclination between the two discs after a flyby encounter can reveal the initial inclination of the encounter, as long as the perturber did not have a disc initially. Here we consider what that means for existing observations of flybys. Observation evidence of a flyby encounter includes: 1) tidally induced spirals, 2) long bridges of material connecting to the perturber, and 3) formation of second-generation discs. Several systems with protoplanetary discs being perturbed by a flyby candidate are currently observed \cite[e.g.,][]{Cuello2023}. The relationship between the disc inclination and the perturber during a flyby encounter is independent of the perturber's mass, periastron distance, or position angle. If a flyby system is identified to have a second-generation disc around the flyby, these steps can be used to reconstruct the tilt of the flyby during the encounter: 
\begin{enumerate}
  \item Measure the mutual inclination between the primary disc and the disc around the flyby candidate. 
  \item If the disc formed during the encounter, its tilt will be roughly twice the initial tilt of the flyby orbit. 
  \item Estimate the tilt of the flyby orbit based on this relationship. 
\end{enumerate}

 The most compelling case to test the relationship between disc inclination and perturber tilt is the system UX Tau. UX Tau is a young quadruple system, located in the Taurus star-forming region. The circumstellar disc around UX Tau A and UX Tac C show signs of dynamical interaction, where the large spirals are detected in the disc around UX Tau A and a long bridge of material extends between UX Tau A and UX Tau C. The rotational signature of the two discs can be clearly seen in the map of the peak intensity velocity \cite[e.g.,][]{Menard2020}. From the observations, the two discs have a mutual inclination of $\sim 80^\circ$ \cite[e.g.,][]{Francis2020,Menard2020}. The disc around the flyby candidate UX Tau C does not show millimeter emission in the disc and the mm-sized dust disc around UX Tau A is more compact than the gas disc. The observations are consistent with the disc around UX Tau C was formed during the flyby encounter. In such a scenario, we can use the results in this work to reconstruct the initial tilt of UX Tau C to be $\sim 40^\circ$.


It is not clear whether the remaining observations of discs around flyby candidates were formed during the encounter, i.e. a second-generation disc, or if the discs around the perturber were present before the encounter. Observations of SR 24 (also known as HBC 262) show a bridge of material between SR 24N and connecting to the disc around SR 24S \citep{Mayama2010,Mayama2020, Weber+2023}, suggesting a flyby event has recently occurred. AS 205 is a triple star system where two components are resolved, AS 205 N and AS 205 S. The discs around each component are misaligned to one another with a bridge of gas between the two sources detected by the ALMA $^{12}$CO (J=2-1) data \citep{Kurtovic2018} and by SPHERE in scattered light \citep{Weber+2023}. From the ALMA observations, the disc around the flyby candidate, AS 205 S, displays millimeter emission, which suggests the disc was present before the encounter. The gaseous bridge between FU Ori N and FU Ori S is misaligned with respect to the disc mid-plane \citep{Perez2020, Weber+2023}, which is evidence of an inclined flyby encounter. An inclined flyby has also been proposed to explain the disc morphology for two systems, Z CMa \citep{Dong2022} and Sgr C \citep{Lu2022}. Further observations are needed to identify whether the discs around the perturber are thought to be second-generation or present prior to the encounter.

\section{Summary}
\label{sec::summary}
We investigated  the interaction of a protoplanetary disc with a grazing parabolic orbit flyby using both $N$--body and three-dimensional SPH simulations. Our simulations and the corresponding analysis were conducted to examine the relationship between the perturber's tilt and the resulting tilt of the second-generation discs. Through systematic variation of the perturber tilt, it was discovered that the tilt of the resulting second-generation discs consistently maintained a proportional relationship, precisely twice that of the perturber.

Through $N$--body simulations, we find a prograde encounter can efficiently capture material when the flyby's periastron is close to the outer disc edge. The captured material can form a second-generation disc around the flyby \citep{Clarke1993,Munoz2015,Cuello2019}. We investigate the inclination distribution of captured particles based on the initial tilt of the flyby orbit. We find a relationship where particles  are captured with a tilt twice the perturber's initial tilt. This relationship is evident in Fig. 20 from \cite{Jikova2016}.

We then consider highly-resolved hydrodynamical simulations of a flyby encountering a protoplanetary disc. We find that the captured, second-generation disc forms at a tilt twice the initial flyby tilt. This relationship holds when we vary the flyby's tilt, position angle, periastron, and mass. Analyzing the disc characteristics, such as eccentricity and tilt, of these second-generation discs can give information about the orbital properties of the flyby encounter \citep{Jikova2016}. Therefore, knowing the relationship between the tilt of the second-generation disc and the tilt of the flyby orbit can be used to reconstruct the trajectory of the flyby provided that there was no disc prior to the encounter. We also simulate the case where the flyby has a disc of material prior to the encounter, and find that the tilt of the eventual circum-secondary disc after the flyby is determined by both the initial state of the circum-secondary disc and the flyby geometry.


The findings in this work carries significant implications for our understanding of disc formation and orbital dynamics. It suggests a robust correlation between the perturber's tilt and the subsequent tilt of second-generation discs, providing valuable insights into the mechanisms governing their formation. Additionally, this observation highlights the importance of considering the relative angular orientations when studying the evolution and characteristics of second-generation discs.

\section*{Acknowledgements}
I would like to thank the referee and editor for their invaluable contributions in enhancing the quality of the manuscript. The authors would like to thank Grant Kennedy and Dimitri Veras for discussions. JLS acknowledges funding from the ASIAA Distinguished Postdoctoral Fellowship. RN acknowledges funding from UKRI/EPSRC through a Stephen Hawking Fellowship (EP/T017287/1). This research was supported by the Munich Institute for Astro-, Particle and BioPhysics (MIAPbP) which is funded by the Deutsche Forschungsgemeinschaft (DFG, German Research Foundation) under Germany´s Excellence Strategy -- EXC-2094 -- 390783311.
This research was funded, in part, by ANR (Agence Nationale de la Recherche) of France under contract number ANR-22-ERCS-0002-01.
This project has received funding from the European Research Council (ERC) under the European Union Horizon Europe programme (grant agreement No. 101042275, project Stellar-MADE). R.D. is supported by the Natural Sciences and Engineering Research Council of Canada (NSERC) and the Alfred P. Sloan Foundation. R.A.B is supported by a Royal Society University Research Fellowship.

\section*{Data Availability}

 The data supporting the plots within this article are available on reasonable request to the corresponding author. The $N$--body integrations in this work made use of the {\sc rebound} code which can be downloaded freely at \url{http://github.com/hannorein/rebound}. A public version of the {\sc phantom} and {\sc splash} codes are available at \url{https://github.com/danieljprice/phantom} and \url{http://users.monash.edu.au/~dprice/splash/download.html}, respectively.


\bibliographystyle{mnras}
\bibliography{ref.bib} 




\appendix
\section{Resolution study}
\label{appendix::resolution}
In Section~\ref{sec::inclined_flyby}, we see that a perturber on a $45^\circ$ inclined orbit forms a $90^\circ$ inclined protoplanetary disc.  Here, we test the resolution to see whether the disc misalignment is robust at higher resolutions. The higher resolution simulation has $4\times 10^6$ particles,  eight times more particles than the lower-resolution simulations, which constitutes a two-fold increase in resolution. 

\begin{figure}
\centering
\includegraphics[width=1\columnwidth]{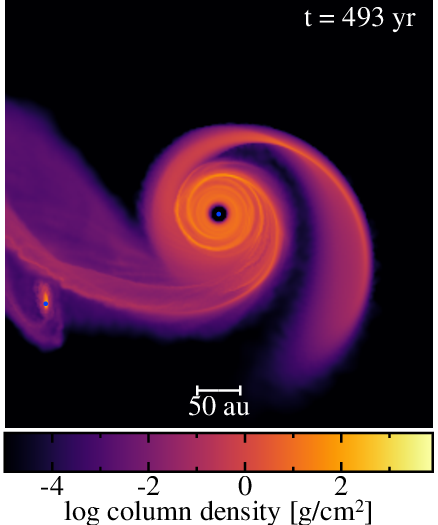}
\centering
\caption{The formation of a protoplanetary discs around perturber during a $45^\circ$-inclined prograde encounter with high resolution (model H45HR). The frame is centered on the primary star, and viewed in the  $x$--$y$ plane, which is face-on to the primary disc. The image is taken at the same time as the lower resolution image in the top right panel in Fig.~\ref{fig::inclined_splash}. The color denotes the disc surface density.}
\label{fig::resolution1}
\end{figure}

Figure~\ref{fig::resolution1} shows the disc surface density for the primary and the perturber discs at a time shortly after the periastron passage of the flyby for the higher resolution simulation. This image is taken at the same time as the lower resolution image in the top right panel in Fig.~\ref{fig::inclined_splash}. The streams accreting onto the perturber disc in the higher resolution simulation are smoother than in the lower-resolved simulation. 

\begin{figure} 
\centering
\includegraphics[width=\columnwidth]{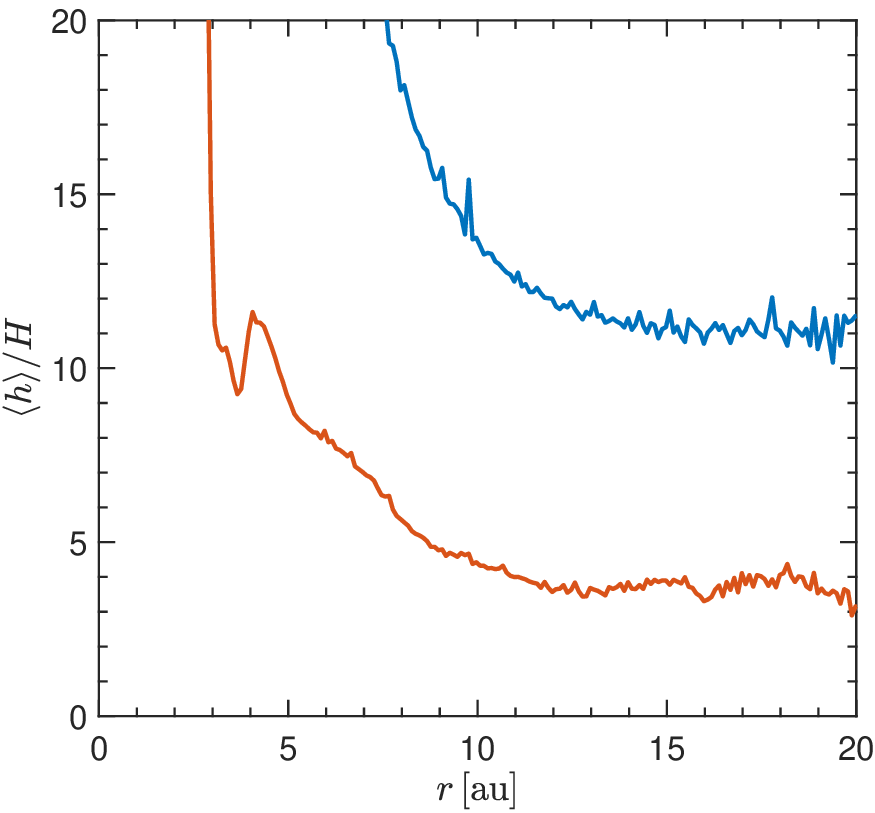}
\centering
\caption{The shell-averaged smoothing length per scale height, $\langle h \rangle/H$, as a function of disc radius, $R$, for the perturber disc at a time of $\sim 5000\, \rm yr$. The blue curve represents the low-resolution simulation ($500,000$ particles, model H45), and the red curve denotes the high-resolution simulation ($4\times 10^6$ particles, model H45HR).  }
\label{fig::resolution_h_H}
\end{figure}

An important parameter that monitors how resolved discs are is the shell-averaged smoothing length per scale height, $\langle h \rangle / H$. Figure~\ref{fig::resolution_h_H} shows $\langle h \rangle / H$ as a function of the perturber disc radius at a time $t = 5000\, \rm yr$. At this time, the perturber disc has damped to twice the initial perturber tilt, which is $90^\circ$ with respect to the tilt of the primary disc. The blue curve shows the $\langle h \rangle / H$ for the lower resolution simulation, and the red curve shows the $\langle h \rangle / H$ for the higher resolution simulation. For the higher resolution simulation, the forming disc around the flyby has an overall lower $\langle h \rangle / H$. However, the disc formed in our high resolution simulation is still unresolved since $\langle h \rangle / H$ is still greater than unity. To reach a $\langle h \rangle / H$ value below unity would require roughly thirty-six times more particles than the higher-resolution simulation, which is beyond our computational resources.



\section{Phase angle}
\label{appendix::phase_angle}
In order to achieve a comprehensive characterization of the three-dimensional orientation of the second-generation disc, two angular parameters are essential: the tilt ($i$) and the longitude of the ascending node ($\phi$). This analysis focuses specifically on the $\phi$ value in each simulation. Notably, simulations featuring a flyby position angle of zero give rise to second-generation discs that exhibit similar $\phi$ values. While the $\phi$ parameter remains relatively stable across these simulations, the disc tilt undergoes changes when the flyby tilt is varied. Consequently, the disc tilt proves to be a more valuable parameter for accurately describing the orientation of the flyby orbit.

\begin{figure*} 
\includegraphics[width=2\columnwidth]{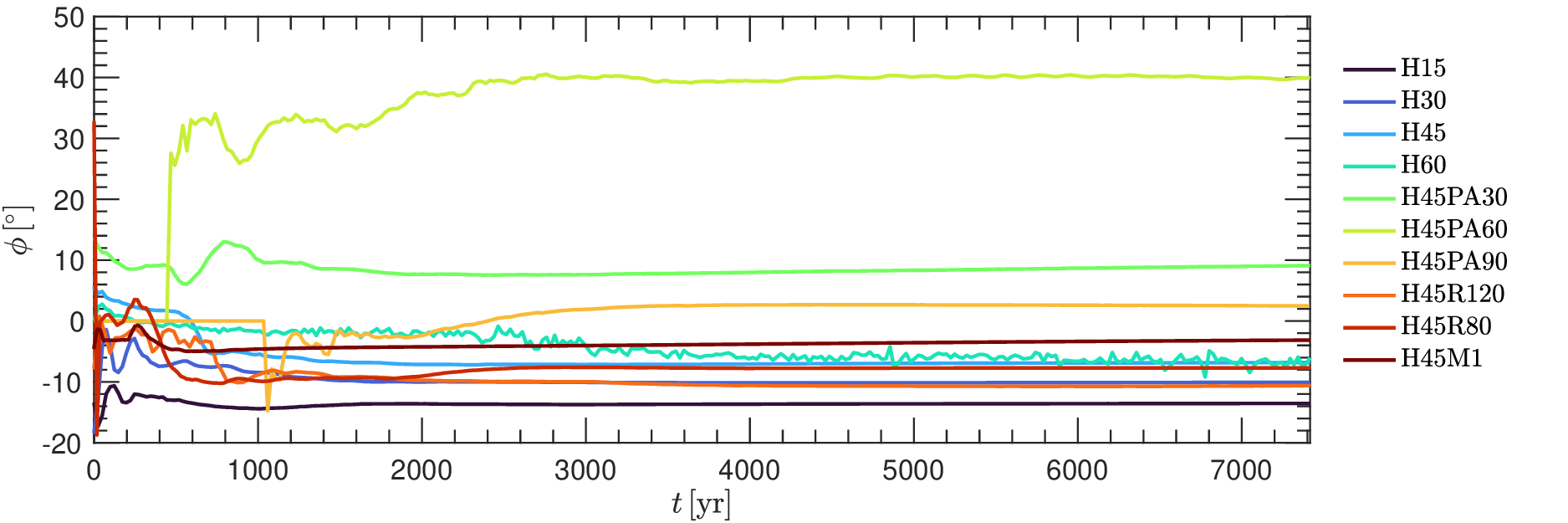}\centering
\caption{ The evolution of the longitude of the ascending node, $\phi$, of the disc forming around the perturber for each simulation parameter (given in Table~\ref{table::setup}) as a function of time. A time $t = 0\, \rm yr$ represents the time of periastron passage.}
\label{fig::phase}
\end{figure*}

\section{Velocity impact calculations}
\label{appendix::analytics}
We calculate the velocity perturbations from \citet{DOnghia2010} using their equations 96 - 107. Here
\begin{align}
    \Delta v_x &= - \frac{2GM_2}{B^2 V_0}r
    \{ [ 2\cos\phi_0 - 3A_x] I_{20}(\sqrt{2}\alpha)  - 3B_x I_{22}(\sqrt{2}\alpha) \nonumber \\ & \quad - 3C_x I_{2-2}(\sqrt{2}\alpha) \}, \nonumber \\
    \Delta v_y &= - \frac{2GM_2}{B^2 V_0}r
    \{ [ 2\sin\phi_0 - 3A_y] I_{20}(\sqrt{2}\alpha) - 3B_y I_{22}(\sqrt{2}\alpha) \nonumber \\ & \quad - 3C_y I_{2-2}(\sqrt{2}\alpha) \}, \nonumber \\
    \Delta v_z &= - \frac{2GM_2}{B^2 V_0}r
    \{ -3A_z I_{20}(\sqrt{2}\alpha) - 3B_z I_{22}(\sqrt{2}\alpha)\nonumber \\ & \quad - 3C_z I_{2-2}(\sqrt{2}\alpha)\}.
    \label{equation:velocity_perturbations}
\end{align}

The generalised Airy functions used in Equations~\ref{equation:velocity_perturbations} are defined in Equations 61-62 and A1-A5 of \citet{DOnghia2010}. The terms $A_x, B_x, ..., C_z$ are themselves functions of the elements of the rotation matrix used for inclined orbits \citep[Section 4,][]{DOnghia2010}. For our problem, with a rotation of $90^{\circ}$ around the $z$ axis followed by $\theta$ around the $y$ axis, the rotation matrix reduces to
\begin{equation}
  \tilde{A} = \begin{bmatrix}
    0 & -\cos\theta & \sin\theta \\
    1 & 0 & 0 \\
    0 & \sin\theta & \cos\theta
\end{bmatrix}.
\end{equation}

Thus the above constants are transformed to
\begin{align}
    A_x &= \cos^2\theta \cos\phi_0, \nonumber \\
    B_x &= 0.5(\cos\theta - \cos^2\theta) \cos\phi_0, \nonumber \\
    C_x &= 0.5(-\cos\theta - \cos^2\theta) \cos\phi_0, \nonumber \\
    A_y &= \sin\phi_0, \nonumber \\
    B_y &= 0.5(1 - \cos\theta) \sin\phi_0, \nonumber \\
    C_y &= 0.5(1 + \cos\theta) \sin\phi_0, \nonumber \\
    A_z &= -\cos\theta \cos\phi_0 \sin\theta, \nonumber \\
    B_z &= 0.5\cos\phi_0 (-\sin\theta + \cos\theta \sin\theta), \nonumber \\
    C_z &= 0.5\cos\phi_0 (\sin\theta + \cos\theta \sin\theta).
\end{align}

Recall here that $\phi_0$ is the phase angle during periapsis passage.


\bsp	
\label{lastpage}
\end{document}